\documentclass[manuscript,screen]{acmart}

\usepackage{multirow}
\usepackage{subcaption}
\usepackage{graphicx}
\usepackage{hyperref}
\usepackage{xcolor}
\usepackage{soul}
\usepackage[normalem]{ulem}
\usepackage[commandnameprefix=always]{changes}

\definechangesauthor[color=BrickRed]{Author}

\usepackage{todonotes}
\setcommentmarkup{\todo[color={authorcolor!20},size=\scriptsize]{#3: #1}}

\makeatletter
\renewcommand\paragraph{\@startsection{paragraph}{4}{\z@}%
                                     {-3.25ex\@plus -1ex \@minus -.2ex}%
                                     {0ex \@plus .2ex}%
                                     {\normalfont\normalsize\textit}}
\makeatother

\AtBeginDocument{%
  \providecommand\BibTeX{{%
    \normalfont B\kern-0.5em{\scshape i\kern-0.25em b}\kern-0.8em\TeX}}}

\setcopyright{acmlicensed}
\copyrightyear{2018}
\acmYear{2018}
\acmDOI{XXXXXXX.XXXXXXX}

\acmConference[Conference acronym 'XX]{Make sure to enter the correct
  conference title from your rights confirmation emai}{June 03--05,
  2018}{Woodstock, NY}
\acmISBN{978-1-4503-XXXX-X/18/06}

\begin{document}

\title[Investigating Conversational Agents to Support Secondary School Students Learning CSP]{Investigating Conversational Agents to Support Secondary School Students Learning Computer Science Principles}

\author{Matthew Frazier}
\affiliation{ 
\institution{University of Delaware}  
\department{Computer and Information Sciences} 
\city{Newark}
\state{DE}
\country{USA}
}
\email{matthew@udel.edu}

\author{Kostadin Damevski}
\affiliation{ 
\institution{Virginia Commonwealth University}
\department{Computer Science} 
\city{Richmond}
\state{VA}
\country{USA}
}
\email{kdamevski@vcu.edu}

\author{Lori Pollock}
\affiliation{ 
\institution{University of Delaware} 
\department{Computer and Information Sciences} 
\city{Newark}
\state{DE}
\country{USA}
}
\email{pollock@udel.edu}

\renewcommand{\shortauthors}{Frazier, et al.}

\begin{abstract}


Secondary school students enrolled in the AP Computer Science Principles (CSP) course commonly utilize web resources (e.g., tutorials, Q\&A sites) to better understand key concepts in the curriculum. The primary obstacle to using these resources is finding information appropriate for the learning task and student's background. In addition to web search, conversational agents are increasingly a viable alternative for CSP students. In this paper, we study the potential of conversational agents to aid secondary school students as they acquire knowledge on CSP concepts. We explore general purpose, generative conversational agents (e.g., ChatGPT) and custom, fixed-response conversational agents built specifically to aid CSP students. We present results from classroom use by 45 high school students in grades 9-11 (ages 14-17) across six CSP sections. Our main contributions are in better understanding how conversational agents can help CSP students and an evaluation of the effectiveness and engagement of different approaches for CSP exploratory search.

\end{abstract}



\received{20 February 2007}
\received[revised]{12 March 2009}
\received[accepted]{5 June 2009}

\maketitle

\section{Introduction}
Research indicates that secondary school students who learn Computer Science Principles (CSP), through one of the curricula developed for this Advanced Placement (AP) course
for learning foundational concepts of Computer Science (CS)~\cite{nsf-ap-csp}, exhibit higher rates of college enrollment and improved skills such as problem-solving \cite{3366808}, response inhibition, planning, and coding \cite{ARFE2020103807}. Online educational resources (e.g., curricula, tutorials, documentation, Q\&A sites) serve as key sources for secondary school students in learning CSP, enhancing classroom learning spanning concept areas including algorithms, security, and the design of the Internet. An impediment to effectively utilizing these resources is the search for information that aligns with the learning task and the learner's background. Research shows that secondary school students need support in searching the web and developing information literacy (i.e., find, evaluate, organize, use, and convey information) \cite{kuiper2005web, KUIPER2009668,doi:10.1207/s15327809jls0901}.
{\em Exploratory search} is particularly challenging as it goes beyond simple lookup and instead is comprised of learning and investigative search intents~\cite{marchionini2006exploratory,asi.23617}.

Research also suggests that conversations are a natural interface for conducting exploratory search~\cite{zhang2018towards, kiesel2018toward, conv_sem_search}. For instance, a growing proportion of difficult web search queries, which do not result in successful information retrieval, are reformulated by users into conversation-starting questions~\cite{aula2010search,liu2012searchfails, 10.1145/3366423.3380193}. The advantages of using conversations with a search system, compared to conventional lookup information retrieval or recommendation systems, are several: 1) question-asking dialogues are a communication modality that humans are well versed in using; 2) the retrieval system can pose clarifying questions to better understand the user need and context; 3) the user can provide incremental positive or negative (explicit) feedback that can help guide the system in the search process.
If designed with customization to build better connections to learning, a conversational agent specializing in exploratory search can also help students relate to the CS concepts by encouraging them to connect CS concepts to their own relevant lived experiences.

This paper focuses on the research question: {\em ``How do fixed-response and generative conversational agents impact the educational process and learning outcomes of secondary school CSP students performing exploratory search tasks?”}.
Conversational agents  (also known as chatbots) can be differentiated by the nature of their responses: {\em fixed-response}, which select from a pre-defined set of responses based on human input, or {\em generative}, which use large language models (LLMs) to generate responses based on the input (or prompt). The fixed-response conversational agents are created with custom responses for a specific task. On the other hand, generative conversational agents are typically used without customization, as general-purpose. While generative conversational agents can technically also be customized (fine-tuned), with, e.g., few-shot prompting, doing so for more than a few inputs is difficult and requires a large dataset and powerful computing resources.

Fixed-response conversational agents are usually built using a set of popular platforms~\cite{9364349} that provide advanced natural language understanding capabilities, e.g., Rasa~\cite{rasa2020}, Google DialogFlow~\cite{dialogflow}, Microsoft LUIS~\cite{luis}, Amazon Lex~\cite{lex}.
Fixed-response domain-specific conversational agents have previously been shown to help support a specified sphere of knowledge. For instance, these agents can help bridge the gap in the instructor's background~\cite{RPAR-STEM-TFAGG,1639593,3162092,10.1145/2716325}. In addition, researchers have shown that in general, secondary school students have high levels of engagement with chatbots~\cite{10.1145/2591708.2591728}. In particular, they could help alleviate social anxiety for students who do not want to ask instructors questions publicly~\cite{10113676,1494907,MP1992CET}.


Generative conversational agents have become increasingly popular due to the recent demonstrated capabilities of LLM-based approaches of OpenAI's ChatGPT~\cite{chatgpt}, Meta's LLAMA~\cite{llama}, Google's Bard~\cite{_2023_bard}, and others. Researchers and educators are still coping with the impact of these agents on learning across the entire educational journey~\cite{baidoo2023education}. 
Generative conversational agents offer the potential advantage of drawing from extensive data sources, enabling them to provide high-quality, pertinent responses to a wide range of student inquiries. However, they also may pose potential drawbacks such as~\cite{chatgpt}: {
\begin{enumerate}
    \item \textit{Factually inaccurate responses}: Sometimes, the generated outputs may sound plausible but are incorrect.
    \item \textit{Unsuitable content for learners}: The responses may not always align with the learner's level or needs.
    \item \textit{Irrelevant information}: The model might include details that are unrelated to the learning task.
    \item \textit{Prompt sensitivity}: The models may require precise prompt engineering, where slight rephrasing can change the output's correctness.
    \item \textit{Lack of clarifying questions}: Despite having the capability, the models often guess user intent rather than seeking clarification.
    \item \textit{Model degradation}: Over time, the quality of responses can deteriorate due to factors such as overfitting during fine-tuning.
    \item \textit{Model collapse}: This refers to a phenomenon where a model's performance drastically declines, often due to repetitive training on similar data.
    \item \textit{Fixed knowledge domain}: The model's understanding is constrained by the data it was trained on, limiting its ability to address new or emerging topics.
    \item \textit{Bias and stereotypes}: The models may reflect biases present in the training data, such as political leanings or cultural perspectives, and may perform best in English. 
\end{enumerate}
}

In contrast, fixed-response, custom conversational agents consistently deliver curriculum-appropriate responses, given that their responses are purposefully crafted for this context. Nonetheless, they may exhibit inflexibility and limit the exploration of various learning pathways.


To answer our research question, we customized a fixed-response conversational agent, called Aida, to a subset of the CSP course. We then evaluated the effectiveness and engagement of Aida, ChatGPT as a representative general purpose, generative conversational agent, and conventional web search using Google. 
Our study consists of two research hypotheses:
(H1) Conversational agents provide better learning experiences than conventional web search for secondary school CSP students performing exploratory search, and
(H2) Generative conversational agents lead to better perceived student experiences despite being less engaging than fixed-response conversational agents.
Both hypotheses were formulated prior to the study. Hypothesis 2 (H2) is based on preliminary observations indicating that the generative AI’s adaptability could offer more effective scaffolding compared to a fixed system. At the time of the study, research directly comparing retrieval-based agents with generative agents is limited, as the latter technology is still emerging~\cite{icml2015, 16435, li-etal-2016-diversity, ritter-etal-2011-data, Davis2018}.

To evaluate our empirical hypotheses, we recruited 45 high school students in grades 9-11 (ages 14-17) to use the agents to complete two tasks on topics within the CSP curriculum. We also gathered perspectives from these students through a post-survey.
Specifically, this paper makes the following contributions to the Computing Education community: 1) a fixed-domain conversational agent for CSP; 2) metrics for the effectiveness and engagement of agents for learning; and 3) a comparative study involving 45 secondary school students using conversational agents and web search in a CSP class.
Our results can help researchers to better understand students' needs for designing future conversational agents for CS learning through exploratory search. 

\section{Related Work}

Created by the National Science Foundation in 2014, CSP is an AP secondary school curriculum framework
for learning foundational concepts of CS~\cite{nsf-ap-csp}. 
More than a traditional introduction to programming course, CSP challenges secondary school students to explore how computing and technology can impact the world by educating students on ethical, inclusive, and collaborative computing culture. 
The CSP framework is grounded in five areas: (1) Creative Development, (2) Data, (3) Algorithms and Programming, (4) Computer Systems and Networks, and (5) Impact of Computing. 
Professional development programs~\cite{11453159496, 10.3603486} and peer assistants~\cite{10.1145/3585077} have been shown to help secondary school CSP instructors cope with the challenges of teaching CSP with little prior CS teaching experience, yet additional support is needed for CSP instructors and students~\cite{10.1145/3617599, 10.1145/3572900}.
Students and instructors in CSP often search the web to acquire knowledge related to the course~\cite{Wang2010,kuiper2005web,KUIPER2009668,doi:10.1207/s15327809jls0901, 3355375}.
Instead of printed manuals and books, they use a variety of online sources including Q\&A forums, CS-related documentation, tutorials, chats, etc~\cite{chatterjee2017what,sadowski2015developers,abdalkareem2017developers}. CS students regularly search these sites to lookup information, reuse code, and learn new concepts and skills. While search is routinely used by software engineers,  support for effective searching for CS-related information is far from a solved problem for a variety of reasons~\cite{liu2020opportunities}. 

Literature on ChatGPT in education or learning context has emerged across many subject areas~\cite{yun2023chatgpt, doi:10.1080/2331186X.2023.2243134} including biochemistry, health professionals, chemistry, business management, nursing, multidisciplinary, social sciences, medicine~\cite{medical}, and computer science~\cite{3587102.3588815, 3587103.3594206}. 
Prather et al.~\cite{prather2023robots} conducted a literature review that explored large language models in computing education.  
They classify LLM literature into five broad categories: 1) assessing the performance, capabilities, and limitations of LLMs (i.e., programming exercises~\cite{10.3588814, piccolo2023bioinformatics, poldrack2023aiassisted, 103603474, kiesler2023large, 3588794, Malinka_2023, 3587102.3588814, 588805}, multiple-choice question answering~\cite{Savelka_2023, Wang_2023}, textbook and exam questions~\cite{Jalil_2023, 103593695}, and assisting the learner through hints or Socratic questioning~\cite{al-hossami-etal-2023-socratic, 3603476, widjojo2023addressing}); 2) using LLMs to generate teaching materials~\cite{sridhar2023harnessing}; 3) using LLMs to analyze student work~\cite{al-hossami-etal-2023-socratic}; 4) studying the interactions between programmers and LLMs~\cite{babe2023studenteval, nam2023inide}; 5) position papers and surveys/interviews~\cite{denny2023computing, brusilovsky_2023_the, Bull_2023, mti7080081, 10.13593349, 10.113600138, ma2023ai, 103610406, 103593360, Yan_2023, zastudil2023generative, doi:102284901}. 
Researchers have assessed ChatGPT's capability of generating code in a variety of contexts~\cite{Codex1, 1123.3576134, ross2023case, 3569823, Li_2022}, e.g.,  Codex GPT-3 LLM, Github Copilot, and Google DeepMind Alphacode. ChatGPT has been shown to easily solve CS1 and CS2 coding problems~\cite{prather2023robots}. 
While many educators are concerned about the threat that ChatGPT imposes to academic integrity and the learning development of their students, one study showed that code generated by ChatGPT had enough differences from student generated code that it could be detected very accurately \cite{idialu_whodunnit}. A study by Hou et al. suggests that although LLMs are being rapidly adopted, they have not yet fully eclipsed traditional help-seeking resources ~\cite{Mettille2024}.

Conversational agents (or chatbots) are increasingly being incorporated into education pedagogy~\cite{Wambsganss2021, weber2021} including students' hybrid web e-learning environments~\cite{9069183, 9532611}, supporting discussion threads~\cite{10.1145/1111449.1111488, 3594202}, providing feedback on programming assessments\cite{103588852}, help requests~\cite{Hellas_2023}, and summarizing Stack Overflow threads~ \cite{answerbot17, 10.1145/3338906.3341186}. Educational researchers have observed that chatbots can assist secondary school CS instructors~\cite{doi:2014.965823}, especially those lacking a CS background~\cite{RPAR-STEM-TFAGG,1639593,3162092,10.1145/2716325}. In addition, researchers have shown that in general, secondary school students have high levels of engagement with conversational agents~\cite{10.1145/2591708.2591728}. Chatbots could help alleviate social anxiety for students who do not want to ask instructors questions publicly~\cite{10113676,1494907,MP1992CET}. 
Research shows that CS reasoning and code explanations created by LLMs are easier to understand than those created by students~\cite{leinonen2023comparing}.
In many aspects, HCI researchers have investigated ways to design and implement conversational agents in educational settings~\cite{10.581247, 10.1145/3544548.3581427, 10.1145/3544548.3580719, 3581318} including prompt engineering~\cite{10.1145/3544548.3581388}.
To our knowledge, no one has studied the effectiveness and student engagement levels of fixed-response versus generative conversational agents for CSP exploratory search tasks.

\section{Evaluating Conversational Agents with CSP Students}
Our study is designed to answer the research question: {\em ``How do fixed-response and generative conversational agents impact the educational process and learning outcomes of secondary school CSP students performing exploratory search tasks?”}  By recruiting students currently enrolled in CSP courses, we compare three approaches to gather information via exploratory searches: two conversational agents and a web search engine. Specifically, through several metrics, we measured the effectiveness of exploratory search for secondary school CSP and student engagement with: (1) Aida, a customized, fixed-response conversational agent, (2) ChatGPT, a general-purpose, generative conversational agent, and (3) Google's web search. 


\subsection{Approaches to Exploratory Search}
\subsubsection{Aida: Fixed-response Conversational Agent}

Since, to our knowledge, there were no existing fixed-response conversational agents for CSP, we designed and built one.
Similar to a flow chart, fixed-response conversational agents map out the conversation and direct the conversational flow using follow-up questions. Such conversational agents contain a retrieval-based response system, i.e., for each user message, the model detects the intent, identifies key entities (e.g., names, locations, dates), and retrieves a predefined response from its repository. Fixed-response conversational agents are also task-driven, which allows us to direct the conversational flow to align with the CSP curriculum. Through follow-up questions, the conversational scaffolding allows us to organize the information being revealed by the agent to a pedagogically sound framework for delivering computer science lessons. 

\begin{figure}[t]
\centering
\subcaptionbox{Aida User Interface\label{fig:aida_ui}}
{\includegraphics[width=0.45\textwidth]
{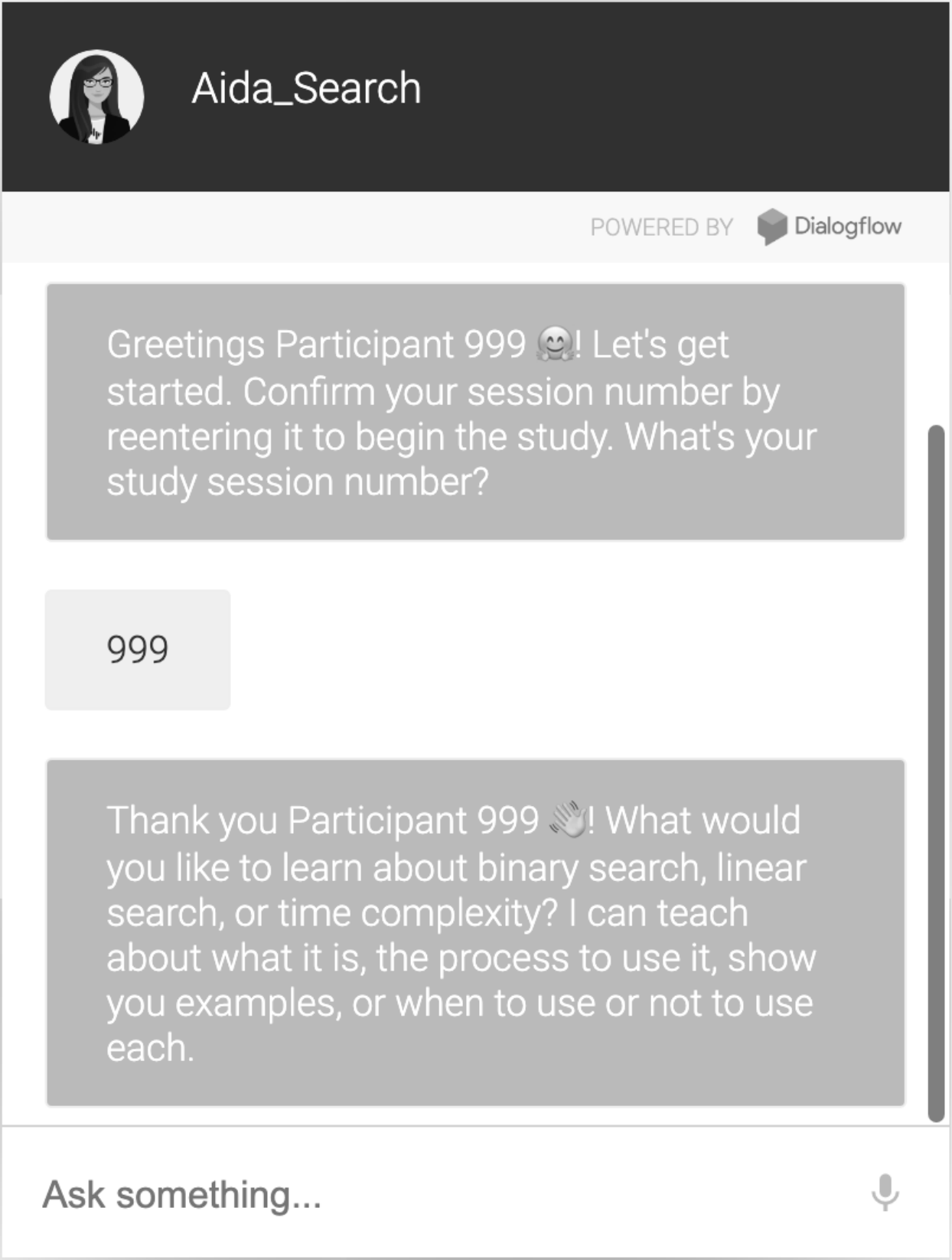}}%
\hfill
\subcaptionbox{ChatGPT User Interface\label{fig:chatgpt_ui}}{\includegraphics[width=0.45\textwidth,height=0.46\textheight]{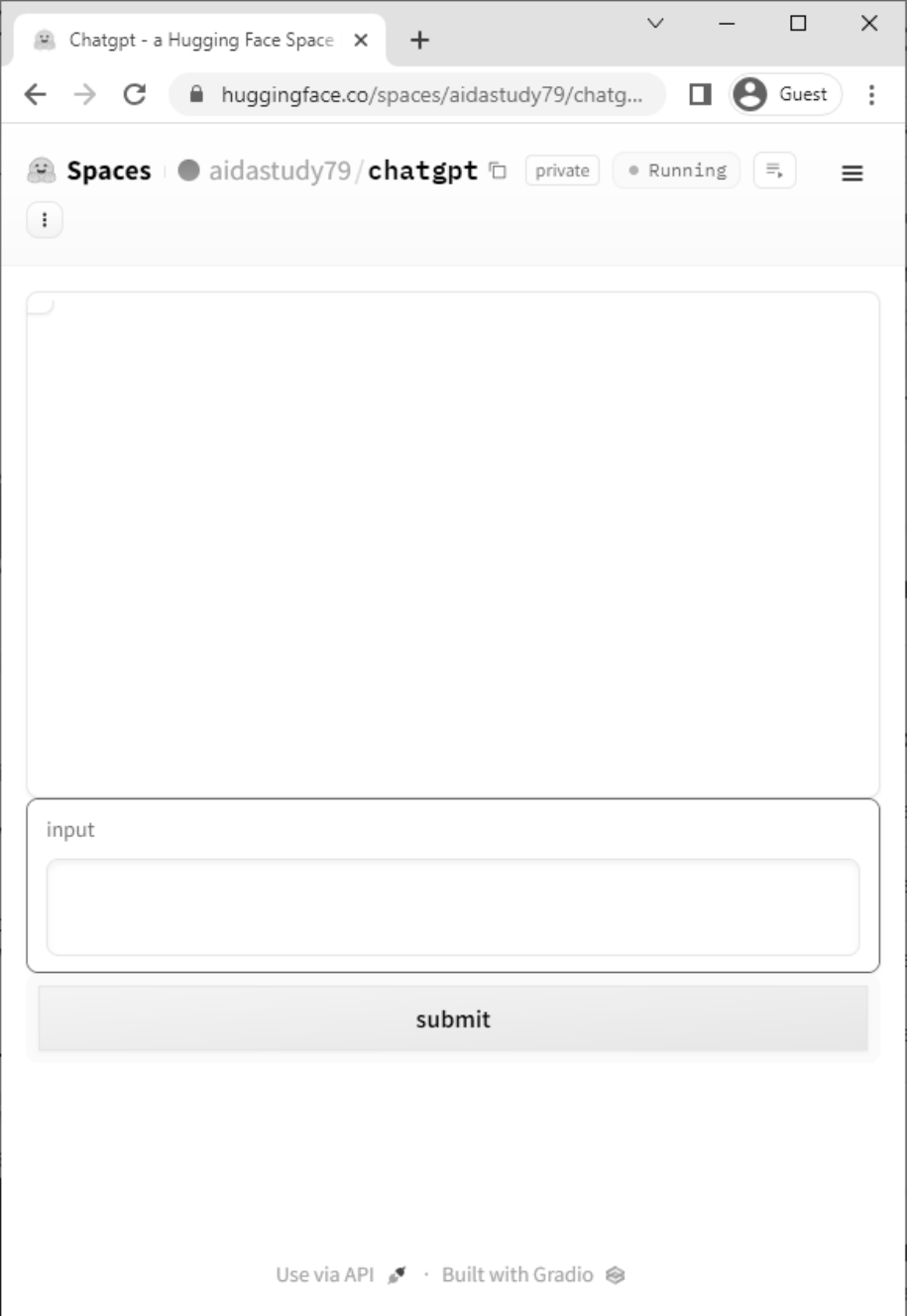}}%
\caption{Conversational user interfaces for with Aida's DialogFlow environment and ChatGPT's HuggingFace environment.}
\label{fig:conversational_interfaces}
\end{figure}

To build the fixed-response agent in our study, Aida (i.e., affectionately named as to ``aid a user''), we chose the Google DialogFlow ES platform~\cite{dialogflow}. Google DialogFlow ES is widely utilized in research~\cite{978-981-99-7353-8_26, Dhotre_Jain_2023, 10.1063/5.0150047} and commercial use~\cite{dialogflow}, as it offers a low barrier to entry, a slot-filling intent strategy, session management, an API interface with code examples in various programming languages, and a simple user interface. Specifically for the demographics and purpose of this study, DialogFlow ES meets the requirements for research in pedagogical conversational agents. Following Weber et al.'s taxonomy, a rule-based back-end technology with a text interface and a persona is an appropriate pedagogical conversational agent design for secondary school students when the role of the agent is to act as a tutor for initial or actual learning of conceptual knowledge in CS~\cite{weber2021}. 

Aida's design aligns with the learner-centered design principles for conversational agents in educational settings, as outlined by Schmitt and Wambsganß et al.~\cite{agent-design-principles1}. More specifically, Aida is adapted to a specific educational setting, is empathetic, and includes an avatar. Aida identifies and addresses the most frequently asked questions (albeit for predefined topics), has appropriate answers, and offers further useful information for students. 
At the time of the study, Aida was not deployed in each school's identity yet; however, Aida was easily accessible through a web-based application and convenient to use. 
To respond to student messages, Aida utilizes both rule-based grammar matching and machine learning (ML) intent matching (classification threshold of 0.3). Students accessed Aida's user interface shown in Figure~\ref{fig:aida_ui} through one of two custom URLs depending on their assigned task for the study: (1) \href{https://bot.dialogflow.com/5b65b258-b361-4b80-ac73-f25372d03b2f}{Binary/Linear Search task} or (2) \href{https://bot.dialogflow.com/25545f22-2610-4426-9800-0533ae137346}{API/Library task}.

\begin{figure}[t]
    \includegraphics[width=0.95\linewidth]{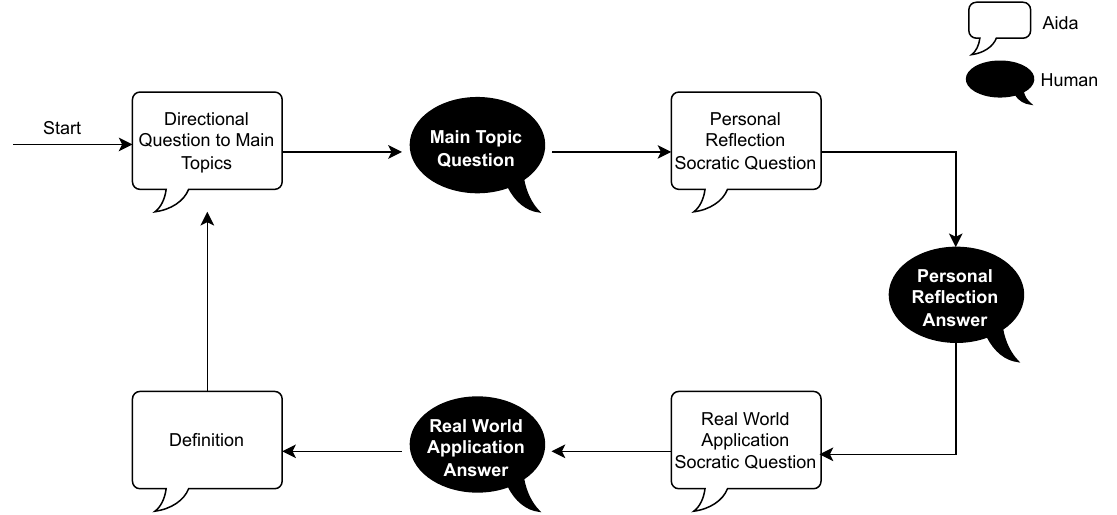}
	\caption{Pedagogical DialogFlow Follow-Up Intent Design. }
	\label{fig:diagram1}
\end{figure}

\paragraph*{Aida: Pedagogical Intent Design Methodology. }

\ Figure~\ref{fig:diagram1} shows the pedagogical design of Aida.
Aida uses the Socratic method as a teaching strategy to allow students to share their thoughts, expand their understanding, and explore new related ideas in the context of their current work~\cite{a2023_ap}. We designed the rule-based intents to align with Code.org's CSP pedagogy~\cite{csp-tps}, which follows Kaddoura's Think-Pair-Share pattern~\cite{kaddoura2013think}, a strategy for enhancing student's critical thinking skills. 
In particular, their tools and materials are designed to support exploration and discovery by those without computer science knowledge, so that students can develop an understanding of these concepts through “play” and experimentation~\cite{codecurriculum}. 
In the Think-Pair-Share pattern, the instructor first asks an open-ended question and students think quietly about it for one to two minutes. Second, students form pairs where they discuss the question for two to five minutes with classmates. Third, the entire class engages in a discussion where each student pair shares their thoughts and ideas they've gathered. Last, the instructor facilitates the student's ideas while summarizing the key points as it relates to the relevant learning objectives. 

\begin{figure}[t]
    \includegraphics[width=0.95\linewidth]{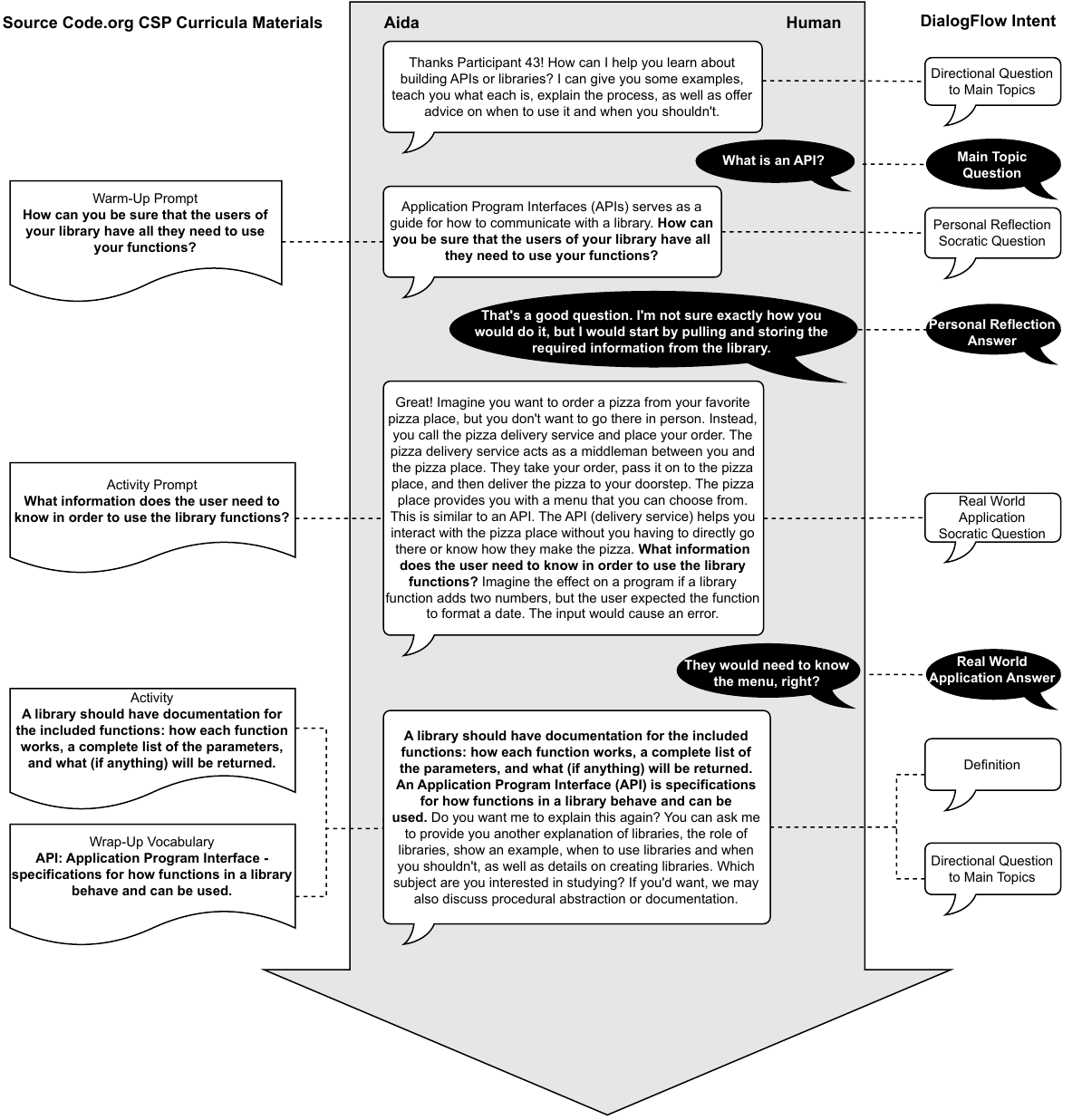}
	\caption{Pedagogical Conversation Example of DialogFlow Follow-Up Intents. }
	\label{fig:diagram2}
\end{figure}

Code.org's CSP lectures parallel the Think-Pair-Share pattern into their own form of a three-part sequential pattern: 1) Warm-Up; 2) Activity; 3) Wrap-Up. 
Figure~\ref{fig:diagram2} illustrates an example conversation where Aida incorporates the Warm-Up, Activity, and Wrap-Up CSP curricula materials into the DialogFlow intent pedagogical design. Instructors first introduce CS concepts to students through an open-ended question to think about for a brief period as a Warm-Up. Next, students join pairs to work on an Activity where they brainstorm and solve questions while sharing their thoughts with their partners. After the activity, all student pairs share their thoughts with the class, and the instructor summarizes the concepts in the lesson in a Wrap-Up. Aida incorporates this design into its conversational responses by assuming the partial role of the instructor and student in our intent design. For primary CS concepts related to the learning task, Aida's initial response ends in an open-ended question to elicit critical thinking from the student as a Warm-Up. Next, Aida's responses simulate interactions of pairing with a partner and sharing ideas through conversation by offering real-world examples and fallback phrases related to the open-ended Warm-Up question. Last, Aida summarizes by giving a formal definition of the primary CS concept to Wrap-Up discussion on a topic. To mimic the Think-Pair-Share pattern in Aida, we build follow-up questions in intents to direct the conversational flow to advance the student through each phase of the Think-Pair-Share pattern. The open-ended follow-up questions are designed to probe the learner, through clarifying assumptions or implications, to come up with the answer on their own.

In addition to following the pedagogical approach in CSP lectures, we provided intents that show examples of processes on CS concepts. To maintain the scope of the knowledge domain, we provide definitions for related CS concepts. For any non-related CS topic asked to the model, we customized a response to redirect the student's attention back to the topics relevant to their task.

\paragraph*{Aida: Intent Content. }\ A key component of Aida's fixed-response design is conveying content that is appropriate for a secondary school student's learning level. Given that Aida is a fixed-response agent, the chances of receiving the same response more than once are higher for a given query and thus, it is imperative to curate many responses to a wide variety of inputs such that Aida appears more human-like.  To diversify Aida's responses, we aggregated text data from two CSP curricula consisting of: 1) Code.org's CSP lecture slides; 2) Code.org's CSP student web modules; 3) The Beauty and Joy of Computing instructional videos; and 4) The Beauty and Joy of Computing CSP student web modules. We provide each intent with a minimum of 20 training phrases to anticipate a variety of student inputs. Each intent has a minimum of 6 different responses where each response is grounded in different CSP course curricula material. Aida contains 69 intents in total across both tasks in our study (35 intents for Binary/Linear Search Task and 34 intents for API/Library Task).

\subsubsection{ChatGPT: Generative Conversational Agent}
For the general-purpose, generative conversational agent, we used OpenAI's ChatGPT ({\em gpt-3.5-turbo}). We chose ChatGPT for our generative conversational agent because at the time of writing it was the most popular conversational agent based on large language models. We aim to simulate a real-world experience that students would encounter using ChatGPT on their own. We chose the GPT-3.5-turbo model that powers the free version of ChatGPT which many students are likely to use.
The model temperature for ChatGPT's web interface is 0.7, and we chose 0.7 as the model temperature in our agent. We explicitly considered but decided not to adjust the model temperature in our design. We intentionally chose to reproduce an experience similar to what a student would encounter on the web. Further, to emulate the ChatGPT web interface, we chose not to manipulate the API system prompt or user's input prompt into the model in any way. We acknowledge that there may be a custom system prompt in the web interface of ChatGPT, however, we were unable to find any resources that support this claim. Further, nothing prohibited students from implementing prompt engineering, as would be required of them when using ChatGPT's web interface on their own, where a custom system prompt could be initiated by the students.
%
%
Students communicated with ChatGPT through a HuggingFace~\cite{huggingface_2023_hugging} environment created specifically for each student in this study.  
HuggingFace is a popular free platform widely used in research that comes pre-built with common machine learning models, datasets, and applications readily available. We chose HuggingFace to host ChatGPT for its ease of set-up, reliable security, high availability of the platform, and to minimize costs. The user interface contains three components as shown in Figure~\ref{fig:chatgpt_ui}: (1) a conversation history log, (2) a text box to curate their input (brainstorm questions), and (3) a submit button to send messages to OpenAI’s ChatCompletion API~\cite{_2023_openai_api}. Using a custom interface for ChatGPT instead of general  site for this tool allowed us to directly collect the student interactions. 

\subsubsection{Conventional Web Search}
We chose Google as the  search engine for our study because it is the world’s largest and most widely used search engine. Research shows that students search the internet more frequently than they ask peers for help, and ask peers for help significantly more frequently than they ask instructors for help~\cite{pcssohet, prather2023robots}. 




\subsection{Pilot Sessions}

Prior to the main study, we conducted five pilot sessions with students who did not participate in the study. Four sessions used Aida and one used ChatGPT to evaluate their functionality and alignment with the CSP pedagogical framework. Feedback from these sessions led to iterative improvements in Aida's design, including adjusting the tool to facilitate better conversational flow. These pilot sessions also helped us refine the user interfaces for all tools, ensuring the tasks were appropriate for secondary school students, and also helped us adjust our classroom study procedure.

\subsection{Participants}

We recruited three high school instructors (each from a different secondary school in a region of the northeastern United States) to conduct our study with the students in their AP CSP classrooms. Each instructor teaches two sections of the course where each section has 15-20 students. 
Not all students from every section participated in the study; only students who obtained parental consent and child assent prior to the start of the study participated.
The AP CSP course contains the same content across all sections, yet depending on the pace of the  learning experience, each section may be at a slightly different point within the curriculum. All instructors were willing to have their students participate in our study during one class period. Each instructor reviewed our study protocol, including the purpose of the study, an explanation of exploratory search tasks, and the specific tasks their students would accomplish during the study.

Of the 120 students across the six AP CSP course sections, we conducted the study with 45 high school students who obtained parental consent and child assent to participate in our study. 
Table~\ref{tab:demographics} shows the student demographics by gender, grade, race or ethnicity, and experience level.
Students ranged in school year grades 9-11 and had varying levels of prior introductory computer science experience except two students with no programming experience.
Each instructor assisted in recruiting the students for the study by facilitating advertisements through an email to the parents of the students. 

\begin{table}[h]
\caption{Student Participant Demographics by Gender, Grade, Race/Ethnicity, and Experience Level.}
\small
\centering
\vspace{-3mm}
\begin{tabular}
{p{0.1\columnwidth}p{0.4\columnwidth}ccc}
\\ \hline
& & Man & Woman & Non-Binary
\\ \hline\hline
\multirow[t]{4}{0.1\columnwidth}{Grade} & Freshman Grade 9 & 7 & 4 & 0 \\
& Sophomore Grade 10 & 15 & 6 & 1 \\
& Junior Grade 11 & 9 & 3 & 0 \\
\multirow[t]{5}{0.1\columnwidth}{Race/Ethnicity} & American Indian or Alaska Native & 1 & 0 & 0 \\
& Black or African American & 8 & 1 & 1 \\
& Other & 6 & 5 & 0 \\
& White & 16 & 7 & 0 \\
\multirow[t]{6}{0.1\columnwidth}{Experience (All that apply per student)} & Program at home on my own & 9 & 2 & 0 \\
& Have block-based programming experience & 20 & 5 & 0 \\
& Have taken programming courses in middle school & 17 & 9 & 1 \\
& Have taken programming courses in high school & 30 & 13 & 1 \\
& Have programmed in Python or a similar language & 17 & 11 & 0 \\
& No programming experience & 2 & 0 & 0 \\
\end{tabular}
\label{tab:demographics}
\end{table}

\subsection{Task Design}

To engage students in exploratory search activities for learning CSP, we designed tasks that necessitate multiple search activity iterations and interpretation using Athukorala et al.’s approach~\cite{10.1002/asi.23617} to task design. There are many candidate search activities that have been identified in the literature~\cite{marchionini2006exploratory}; however, we focused our study on students completing two tasks in Marchionini’s “Learn” Exploratory Search category: Comparison and Knowledge Acquisition. We chose these tasks because Comparison and Knowledge Acquisition tasks are suitable for searching as a learning process (i.e., multi-step educational search) since they entail critical and receptive cognitive learning through the student’s learning behavior (i.e., recalling, comprehending, contrasting, aggregating, and synthesizing) and search behavior (i.e., acquiring, comparing, and evaluating usefulness)~\cite{rieh2016}. We believe that these tasks are representative of common daily tasks that secondary school CSP students take over long periods of time such as reviewing prior lecture materials and completing homework assignments that minimize gaps of knowledge in CS concepts.

\begin{figure}[t]
\centering
\subcaptionbox{Binary Search}{\includegraphics[width=0.44\textwidth]{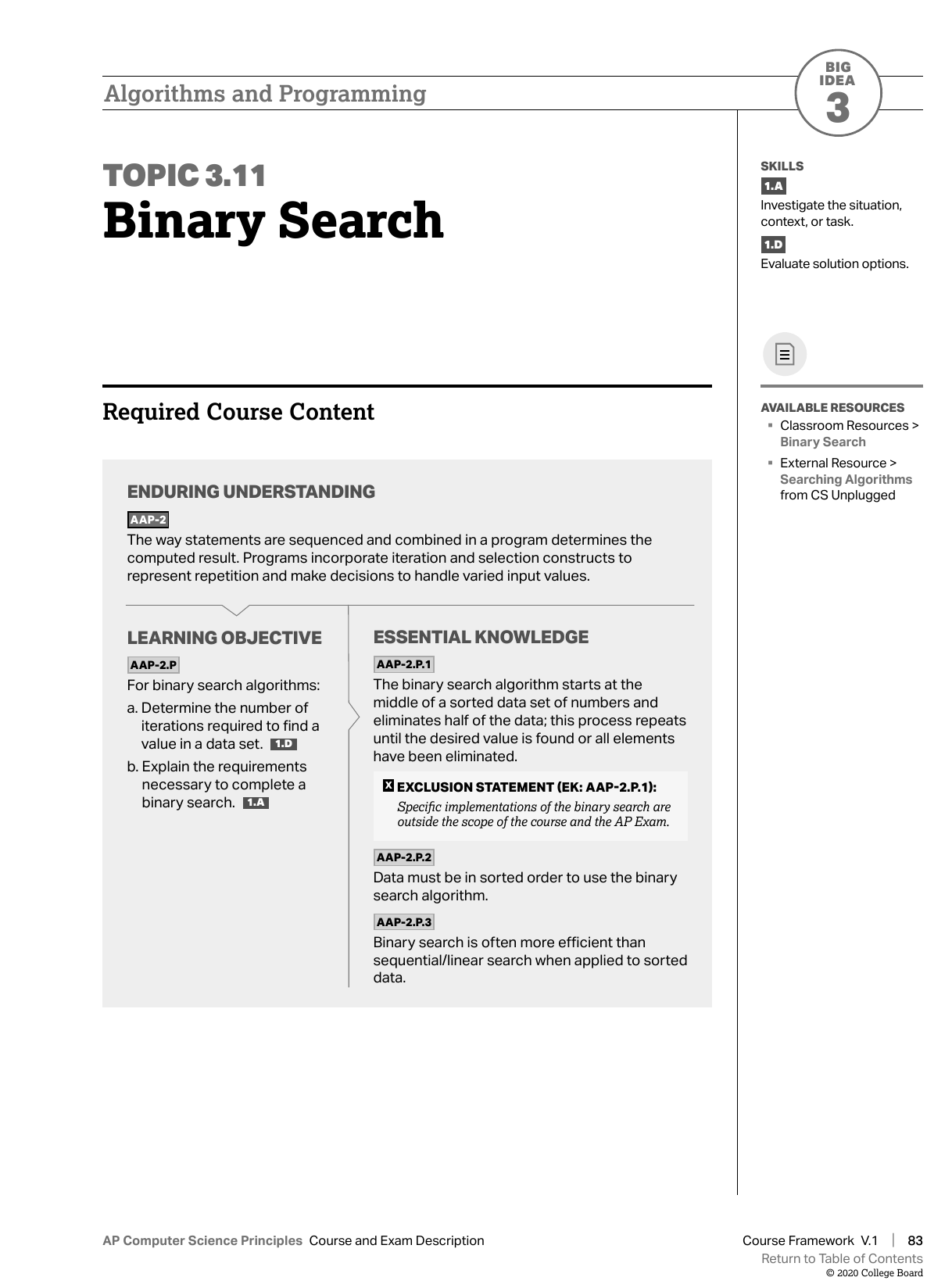}}%
\hfill
\subcaptionbox{Libraries}{\includegraphics[width=0.46\textwidth]{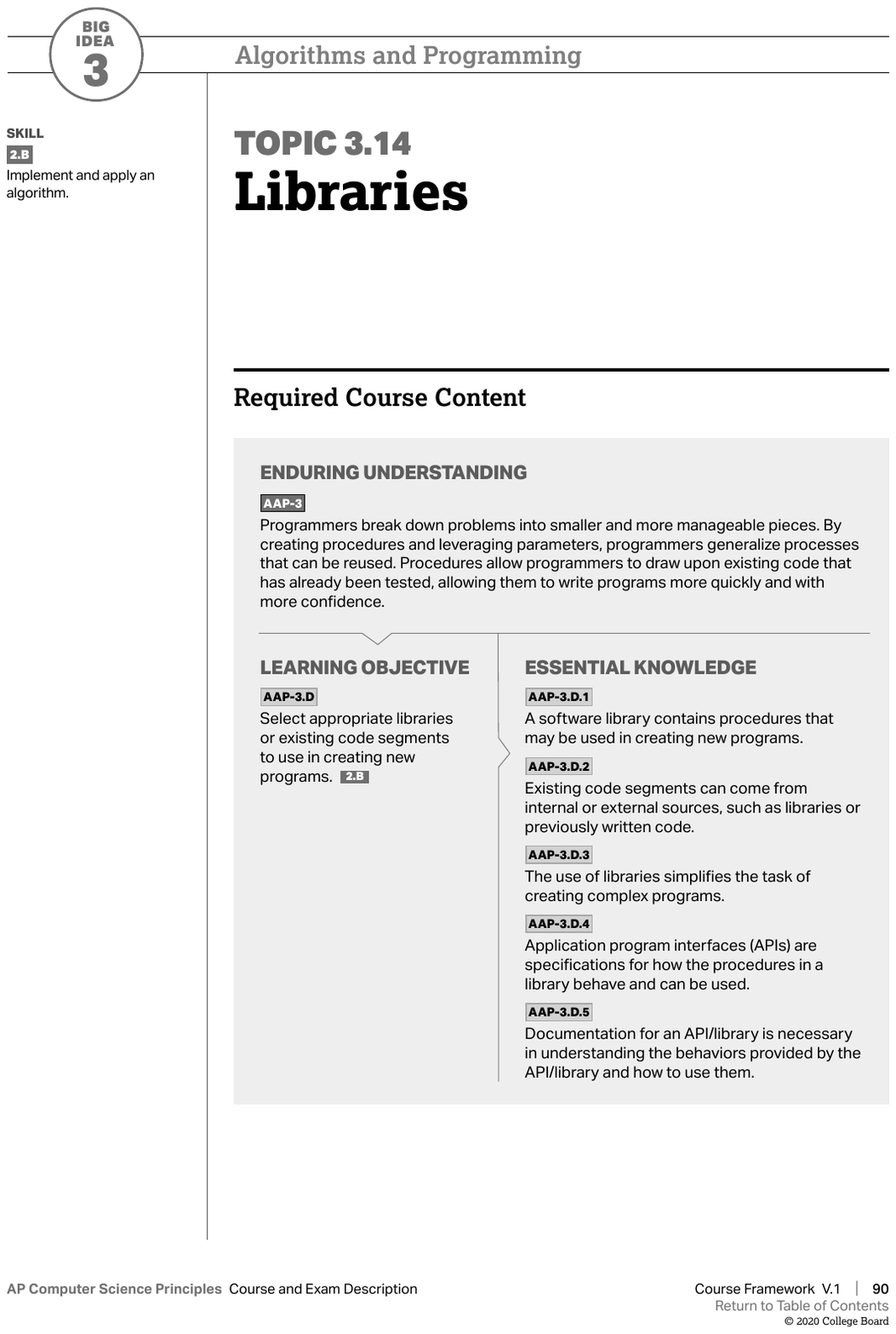}}%
\caption{AP College Board Computer Science Principles Curriculum Framework.}
\label{fig:csp-framework}
\end{figure}

We chose search algorithms and API/libraries as the two  topics for our study because they are more complex topics in the CSP curriculum that motivate exploratory learning.  We created two tasks such that information being explored in each task aligns to the Essential Knowledge in the course’s required content for those topics as shown in Figure~\ref{fig:csp-framework}.
Table~\ref{tab:tasks} shows the task descriptions. The Binary/Linear Search Task involves analyzing the similarities and differences among the binary  and linear search algorithms. This includes comprehension of how each algorithm works, discernment of when to use each algorithm (sorted vs. unsorted lists), and understanding the impacts of each algorithm’s efficiency. The API/Library Task requires students to gain an objective understanding of the requisite components of an effective API/library. This includes comprehension of what APIs/libraries are, how they work, communicating API/library usage through documentation, and understanding that errors may arise due to improper use of APIs/libraries. The expected outcome of each task was for students to write a paragraph summarizing the answer to the question provided, using information gathered through exploratory search during the activity.

\begin{table}[h]
\caption{Exploratory Search Tasks.}
\small
\centering
\begin{tabular}{|p{0.95\columnwidth}|}
\hline
{\bf Binary/Linear Search Task (Comparison):} Imagine that you have been asked to write a program that searches for an item in a list. You have the option to use binary search or linear search. Learn enough to write a short paragraph comparing and contrasting each search algorithm and how each works. What would you have to know about the search problem to know which algorithm to choose?\\ \hline
{\bf API/Library Task (Knowledge Acquisition):} You’re tasked with sharing reusable code from a program that you’ve created with your friends. You will create a library of functions and an API. Learn enough to write a short paragraph on how each works. Write a short paragraph summarizing the key points that you need to keep in mind while you’re creating a library and API. What would you have to know about how your friends intend to use the reusable code from your program to know how to build a library and API? \\ \hline
\end{tabular}
\label{tab:tasks}
\end{table}

\subsection{Classroom Protocol}
The study took place at each secondary school during one CSP class period. 
We designed the study protocol to fit within 50 minutes (the shortest class duration) to accommodate all of the instructor's  schedules.
Each study session consisted of an introduction, sequential completion of the two tasks, and a post-study survey. 
This study employed a within-subject design, where each participant interacted with both Aida and a comparison tool (either Google or ChatGPT) before completing a survey to assess their experiences.
We varied the ordering of the use of the conversational agents and tasks among students to avoid learning bias. We created eight variations of the protocol for each tool/task combination of the study. To minimize risk of introducing unintentional bias, each protocol was randomly distributed to the students. For each protocol, the students were given a different version of the instructions that included the study tasks and directions on how to access the exploratory search tools. 
Each student was allotted up to 15 minutes to complete each exploratory task.

The researcher attended each study field experiment. With the help of each course's instructor, the researcher led the activity and facilitated the study including monitoring the pace and attention of student's progress during the study to ensure successful completion and answer student's questions.
Our study’s instructions were online documents stored on the researcher’s university server. The students accessed the instructions using the web browser on computers in the classroom via a unique short URL. Each instruction set contained space for the students to take notes and curate the outcome of their learning as well as a space to store the history of their session which was collected after completing each task. 

Students were told to not use any handheld devices and close all computer applications while participating in the study with the exception of the web browser. The goal was to avoid any potential distractions as well as allow students the ability to brainstorm and perform notetaking while performing the exploratory search tasks. All students were free to use each tool in any way of their choice. They were not given any instructions on how to interact with each conversational agent, except to avoid jargon. 

\subsection{Post-Study Survey}
We chose not to conduct a pre- and post- study survey of the participant's knowledge of the task (as in traditional comparative analysis) because, given the timing of this field study in May 2023, the topics were likely already covered in the CSP course. Therefore, the comparative knowledge gained from completing a task with each conversational agent may not reflect the true educational impact of the conversational agents.
Instead, we conducted a post-study survey centered around comparing the utility of each of the agents towards completing the tasks.

The post-study survey gathered information on the effectiveness and the interactivity of the conversational agents. 
The survey begins with 5-point Likert scale questions that solicit feedback on the students' perception of how well the conversational agent worked for completing the task including: 1) ratings of how useful the information found by the conversational agent was; 2) if the student perceived that the conversational agent helped them; 3) if the student would use the conversational agent in the future.
The survey also contains 5-point Likert scale questions to obtain feedback on the interactivity of the conversational agent including: 1) the student's ability to understand the conversational agent; 2) the student's perception on if the conversational could understand them; 3) if the conversational agent exhibited or behaved in a non-human manner; 4) the frequency of the conversational agent's follow-up questions.
We also include open-ended questions to collect any additional feedback on what student's like or dislike about each conversational agent and any other comments that they want to provide.
Next, we collect a comparative conversational agent evaluation where students select their preference of each tool relative to: 1) presenting content that relates concepts to personal experience; 2) the method that made the activity easiest to complete; 3) presenting content that is personalized to secondary school terminology; 4) the method that was the most enjoyable as a learner.
Last, we conclude the post-study survey by collecting the student's demographic data.


\subsection{Collected Data}

We collected four key kinds of data from each student’s session: (1) the  conversational agent’s dialogue, including timestamps, (2) the student’s web browser history including the timestamps of the websites visited and Google search queries, (3) the student’s post-survey answers, and (4) the outcomes of each student’s learning experience (the student’s answers to each task).
Aida conversational agent dialogues were collected using the built-in conversation history within the DialogFlow system. Both the ChatGPT dialogues and web browser history were stored in the space provided in each student’s instruction set. 

\section{Data Analysis Methods and Findings}

To answer our research question 
\textit{“How do fixed-response and generative conversational agents impact the educational process and learning outcomes of secondary school CSP students performing exploratory search tasks?”},
we analyzed the data we collected to investigate the effectiveness and user engagement, using several different metrics.  Prior to our statistical analysis, we tested for a normal distribution by visually checking the distribution of our data through Q-Q plots and histograms of each metric. We conducted an ANOVA analysis comparing the students in the three conditions across both tasks for all metrics of effectiveness and user engagement. A summary of our ANOVA analysis is shown in Table~\ref{tab:significance1}. This section describes the metrics and reports the findings; the next section discusses the overall results.

\begin{table}[h]
\centering
\caption{Overview of results (mean and standard deviation (SD)) of student answer completeness, accuracy, extraneousness, task duration, exploratory actions, presented examples, and tool scope for both tasks across the three tools.}
\vspace{-3mm}
\small
\begin{tabular}
{||p{0.14\columnwidth}|p{0.015\columnwidth}|p{0.11\columnwidth}|p{0.07\columnwidth}|p{0.12\columnwidth}|p{0.07\columnwidth}|p{0.1\columnwidth}|p{0.09\columnwidth}|p{0.07\columnwidth}||}
\hline 
\multirow[t]{3}{0.14\columnwidth}{\bf Tool} & \multirow[t]{3}{0.01\columnwidth}{\bf N} & \textbf{Completeness} & \textbf{Accuracy} & \textbf{Extraneousness} & \textbf{Duration} & \textbf{Exploratory Actions} & \textbf{Presented Examples} & \textbf{Tool Scope} \\
& & mean & mean & mean & mean & mean & mean & mean \\
& & (SD) & (SD) & (SD) & (SD) & (SD) & (SD) & (SD) \\
\hline
\multirow[t]{3}{0.14\columnwidth}{{Aida (A)}} & \multirow[t]{3}{0.01\columnwidth}{45} & 0.33 & \textbf{0.86} & \textbf{0.17} & \textbf{11.67} & \textbf{10.56} & \textbf{1.58} & \textbf{0} \\
{} &  & (0.17) & \textbf{(0.2)} & \textbf{(0.26)} & \textbf{(3.81)} & \textbf{(5.05)} & \textbf{(1.32)} & \textbf{(0)} \\
 \hline
\multirow[t]{3}{0.14\columnwidth}{ChatGPT (C)} & \multirow[t]{3}{0.01\columnwidth}{22} & \textbf{0.48} & 0.82 & \textbf{0.17} & 5.77 & 2.91 & 1.36 & 1.59 \\
 &  & \textbf{(0.19)} & (0.18) & \textbf{(0.21)} & (3.78) & (1.32) & (1.97) & (1.74) \\
\hline
\multirow[t]{3}{0.14\columnwidth}{Google (G)} & \multirow[t]{3}{0.01\columnwidth}{23} & 0.38 & 0.78 & 0.27 & 5.48 & 6.78 & 1.17 & 2.39 \\
 &  & (0.19) & (0.3) & (0.32) & (4.01) & (3.37) & (1.7) & (6.35) \\
\hline
\multirow[t]{3}{0.14\columnwidth}{\textbf{Significance}} & \multirow[t]{3}{0.01\columnwidth}{} & p \textless 0.01 & ns & ns & p \textless 0.001 & p \textless 0.001 & ns & p \textless 0.025 \\
\hline
\multirow[t]{3}{0.14\columnwidth}{\textit{Tukey post-hoc test}} & \multirow[t]{3}{0.01\columnwidth}{45} & A-G, G-C &  &  & G-C & A-G, G-C &  & G-C, C-A \\
\hline
\end{tabular}
\label{tab:significance1}
\end{table}

\subsection{Effectiveness}

To evaluate our first hypothesis (H1) that ``\textit{Conversational agents provide better learning experiences than conventional web search for secondary school CSP students performing exploratory search}'', we analyzed the effectiveness of each approach to exploratory search for secondary school CSP across three dimensions, i.e., the effectiveness of: (1) users achieving the task, (2) helping users learn, and (3) customizing to the CSP course scope.
Figure~\ref{fig:effectiveness_of_achieving_the_task} presents our results. 

\subsubsection{Effectiveness of Achieving the Task}

To measure the students' effectiveness in achieving the task, we analyzed the answers that students provide as the outcome of their learning.  We use three measures: \textit{Student Answer Completeness}, \textit{Accuracy}, and \textit{Extraneousness}. 

\paragraph*{Student Answer Completeness.  }

\begin{table}[h]
\caption{Student Answer Completeness Criteria.}
\small
\begin{tabular}{p{0.45\columnwidth}p{0.45\columnwidth}}
\hline 
    \vspace{-1mm}\begin{center}{ \bf Binary / Linear Search Criteria}\end{center}\vspace{-3mm} & \vspace{-1mm}\begin{center}{\bf API / Library Criteria}\end{center}\vspace{-3mm} \\
    \hline\hline
    \begin{enumerate}
    \item Binary search is a way to find an item in a list. 
    \item Binary search at most checks only half of the list. 
    \item Binary search compares middle value in the list with value being searching for, and eliminates half of the list each comparison, depending on if the value is greater or less than the item being searched for.
    \item Linear search is a way to find an item in a list.
    \item Linear search at most checks all items in the list. 
    \item Linear search is sequential and runs from beginning to the end of the list.
    \item Binary search is more efficient than linear search.
    \item  Binary search is better for larger lists.
    \item Linear search is better for smaller lists.
    \item Binary search is only for sorted lists.
    \item Linear search is for lists of any order.
    \item When choosing between both algorithms, consider the order or the list.
    \item When choosing between both algorithms, consider the size of the list.
    \end{enumerate} &
    \begin{enumerate}
    \item APIs are an interface that provides specifications on how to interact with the library. 
    \item An API transfers data between parties. 
    \item Libraries are a collection of reusable functions that perform specific tasks.
    \item A library works by importing it and calling its functions.
    \item Consider testing to prevent library or API misuse or compatibility issues for users.
    \item Documentation communicates to others how to interact with a library or API. 
    \item Consider the formatting of the library or API such as meaningful names, parameters, data types, and return values. 
    \item Consider the programming language that your users want to use.
    \item Consider the specific tasks that users are accomplishing when building a library or API.
    \end{enumerate} \\
\end{tabular}
\label{tab:completecriteria}
\end{table}

\begin{figure}[t]
\centering
\subcaptionbox{Completeness\label{fig:completeness}}{\includegraphics[width=0.33\textwidth]{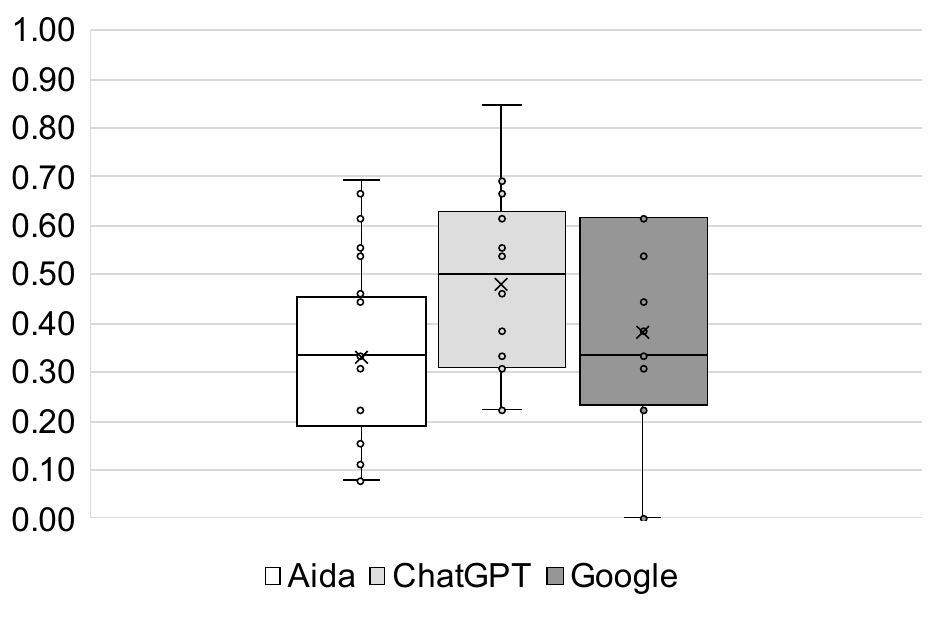}}%
\hfill
\subcaptionbox{Accuracy\label{fig:accuracy}}{\includegraphics[width=0.33\textwidth]{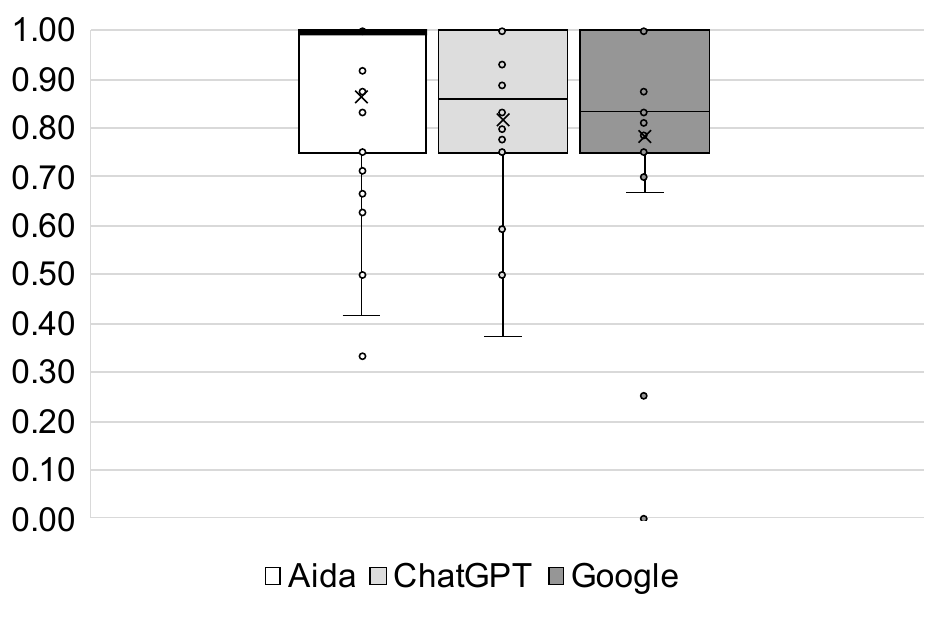}}%
\hfill
\subcaptionbox{Extraneousness\label{fig:extraneousness}}{\includegraphics[width=0.33\textwidth]{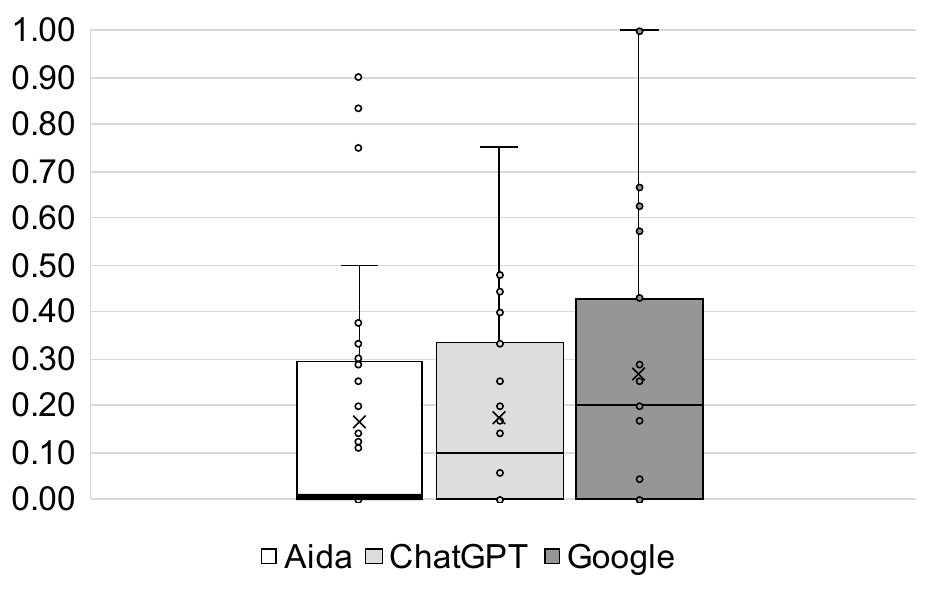}}%
\caption{Distributions of Student Answer Completeness (left), Accuracy (middle), and Extraneousness (right). The results are aggregated across both tasks for each tool. All participants utilized Aida for one task, and either ChatGPT or Google for one task (Aida N=45, ChatGPT N=22, Google N=23).}
\label{fig:effectiveness_of_achieving_the_task}
\end{figure}

\ We define the completeness of the student’s answer as the proportion of information in the student’s answer that we expect to be present in their answer. We consider a student's answer to be ``complete'' if their answer addresses all information items connected to the task that they were assigned. To ensure objectivity and rigor in developing these criteria, three individuals
(one of which is the researcher) with significant CS teaching assistant experience independently generated the expected information items (i.e., the completeness criteria) for each task. These independently developed criteria were then compared and discussed collaboratively to ensure consistency. The researcher merged these sets to create a rubric with the finest granularity of details expected for a complete answer for each task. This collaborative process, involving multiple experienced individuals, was designed to mitigate individual biases and enhance the objectivity of the coding scheme.
Notably, the completeness criteria is intentionally exhaustive by design. We require that the student explicitly states each criterion in their answer instead of relying on a pre- and post- study evaluation of student comprehension. 
Table~\ref{tab:completecriteria} presents the final criteria for completeness of each of our study tasks.
The researcher applied the detailed criteria as a check list on the corpus of student answers.  Each student answer was scored for completeness  as:
\[ 
Completeness=\dfrac{Number\:of\:Completeness\:Criteria\:in\:Student\:Answer}{Total\:Number\:of\:Student\:Answer\:Completeness\:Criteria}
\]

\begin{figure}
    \includegraphics[width=0.99\linewidth,trim={0 22cm 0 2.5cm}, clip]{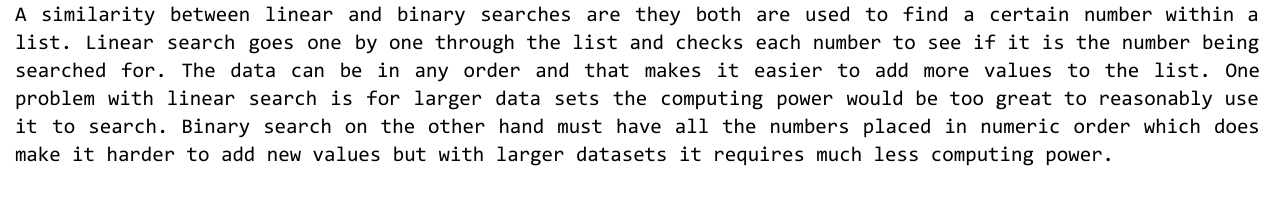}
	\caption{Student Answer Example for Binary/Linear Search Task.}
	\label{fig:output_example1}
\end{figure}

Figure~\ref{fig:output_example1} shows an example of student answers for the Binary/Linear Search task. The student’s answer meets 8/13 criteria for completeness. This student acknowledges that both algorithms find an item in the list of items (Criteria 1 and 4) by stating \textit{``A similarity between linear and binary searches are they both are used to find a certain number within a list''}. This student describes that linear search is sequential (Criterion 6) by stating \textit{“Linear search goes one by one through the list...”} and that the algorithm traverses the entire list for the item being search for (Criterion 5) by continuing that\textit{ “... and checks each number to see if it is the number that is being searched for.''} The student states that linear search is suitable for unsorted lists by answering \textit{“The data can be in any order …“} (Criterion 11) and that binary search is only suitable on sorted lists answering\textit{ “Binary search on the other hand must have all the numbers placed in numeric order…”} (Criterion 10). The student describes that linear search is better suited for smaller lists (Criterion 9) and binary search is more efficient in time complexity than linear search for larger lists in their answer, stating\textit{ “One problem with linear [search] is for larger data sets the computing power would be too great to reasonably use it to search.”} and \textit{“...but with larger datasets it requires much less computing power.”} (Criterion 8). The student’s answer neither states that binary search at most checks half of the list (Criterion 2) nor does the student describe how the binary search algorithm works (Criterion 3). Although the student explains usage on larger lists, the student does not acknowledge that binary search is more efficient overall (Criterion 7). The student does not state that choosing between the two algorithms depends on the size of the list (Criterion 12) and the ordering of the list (Criterion 13).

In Figure~\ref{fig:completeness}, we see that the student answers for those students that used ChatGPT scored higher on completeness than both Aida and Google, which were similar (median of 0.5 for ChatGPT compared to median of 0.33 for both Aida and Google). The small proportions of complete answers, in absolute terms, is attributed only to the detailed and exhaustive nature of our completeness criteria.
Our ANOVA analysis comparing the students in the three conditions regarding their answer's completeness across both tasks is shown in Table~\ref{tab:significance1}. The test revealed that students who interacted with ChatGPT produced significantly more complete student answers compared to both Aida ($M_{Aida} = 0.33, M_{ChatGPT} = 0.478, t = -3.146, Cohen's\:d_{s} = -0.84, p < 0.01$) and Google ($M_{Google} = 0.384, M_{ChatGPT} = 0.478, t = -1.705, Cohen's\:d_{s} = -0.508, p < 0.01$) having a large and medium effect size, respectively.

\paragraph*{Student Answer Accuracy.  }

\ We define the accuracy of the student’s answer as the proportion of expected information items in a student’s answer that are factually accurate. We consider a student’s answer to be “accurate” if the information items in their answer semantically matches the expected information in the student answer completeness criteria for the task being explored. There must be no indication of AI hallucination, misinformation, or inaccurate information from web sources. The researcher assigned a score to each student’s answer representing the ratio of the total count of accurate information items present in their answer out of the total number of expected criteria in the answer. Each student answer received a score, computed as:
\[ 
Accuracy=\dfrac{Number\:of\:Accurate\:Completeness\:Criteria\:in\:Student\:Answer}{Number\:of\:Completeness\:Criteria\:in\:Student\:Answer}
\] 

\begin{figure}[t]
    \includegraphics[width=0.99\linewidth,trim={0 22cm 0 2.5cm}, clip]{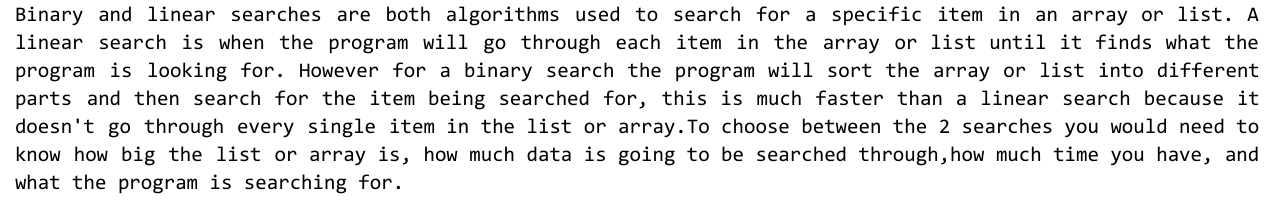}
	\caption{Example of Student Answer containing inaccurate information for the Binary/Linear Search Task.}
	\label{fig:output_example2}
\end{figure}

As noted earlier, the student’s answer in Figure~\ref{fig:output_example1} meets 8/13 criteria for completeness: Criteria 1, 4-6, and 8-11. Of the eight criteria completed in the student’s answer, the student’s answer meets 8/8 in accuracy. This student accurately states that both algorithms are for searching for items (Criteria 1 and 4) by stating \textit{``A similarity between linear and binary searches are they both are used to find a certain number in a list.”}. The student accurately answers that the linear search is sequential and traverses the whole list by stating \textit{“Linear search goes one by one through the list and checks each number to see if it is the number that is being searched for.”} (Criteria 5 and 6). The student accurately answers that linear search is for unsorted lists and that binary search is for sorted lists by stating\textit{ “The data can be in any order …”} (Criterion 11) and \textit{“Binary search on the other hand must have all the numbers placed in numeric order…”} (Criterion 10). The student accurately states that linear search is better for smaller lists (Criterion 9) and binary search is more efficient for larger lists by answering \textit{“One problem with linear [search] is for larger data sets the computing power would be too great to reasonably use it to search.”} and \textit{“...but with larger datasets it requires much less computing power”} (Criterion 8). Figure~\ref{fig:output_example2} contains an example of inaccurate information in a student's answer related to the binary search algorithm. This student inaccurately states that the binary search algorithm involves a sorting algorithm as part of binary search process as opposed to a sorted list being a requirement for utilizing binary search by writing ``\textit{However for a binary search the program will sort the array or list into different parts and then search for the item being searched for...}''. The student inaccurately meets Criterion 3 of the Student Answer Completeness Criteria by including a search algorithm in the divide and conquer methodology of the binary search algorithm.

In Figure~\ref{fig:accuracy}, we see that student answer accuracy is overall high among the three tools. Students using Aida produced slightly more accurate answers. Aida's median accuracy was 1.0 compared to 0.86 for ChatGPT and 0.83 for Google. Our ANOVA analysis ($p = 0.368, p > 0.1$) showed no significant difference in student answer accuracy across both tasks in the three conditions as shown in Table~\ref{tab:significance1}.

\paragraph*{Student Answer Extraneousness.  }

\ We consider a student’s answer to contain ``extraneous” information  if the student provides extra information in their answer that is unrelated to the task being explored, as  compared to the student answer completeness criteria for the task being explored. Supporting information that justify and demonstrate the student’s comprehension of the task being explored, such as examples, are considered to be relevant to the student answer completeness criteria. The researcher assigned a score to each student’s answer representing the ratio of the total count of extraneous sentences present in their answer out of the total number of sentences in the student’s answer. Each student answer received a score, computed as:

\[ 
Extraneousness=\dfrac{Number\:of\:Sentences\:Beyond\:Answering\:the\:Task\:Prompt\:in\:the\:Student\:Answer}{Total\:Number\:of\:Sentences\:in\:the\:Student\:Answer}
\]

Of the extraneous information in the student’s answer, the extraneous text could be reflection, copy-paste facts from the tool's response, or paraphrased facts.  To gain more insight,  we also characterize the text in the student’s answer beyond the text that semantically achieves the completeness criteria. We count the number of student answers that contain each kind of text and then compute the proportion of extraneous sentences  in each session of each kind. 

In Figure~\ref{fig:output_example1}, the student’s answer meets 8/13 criteria for completeness and 8/8 in accuracy, however, the student provides 1/5 extraneous sentences. Along with the student answering that linear search can be performed on unsorted lists, the student provides the first extraneous information item on the difficulty of inserting items into an unsorted list by stating \textit{“...and that makes it easier to add more values to the list”} (0.5 sentence). Additionally, along with the student answering that binary search must be performed on sorted lists, the student provides the second extraneous information item on the difficulty of inserting items into a sorted list by stating \textit{“...which does make it harder to add new values”} (0.5 sentence). We consider both items of information unrelated to the task being explored because the task is to compare and contrast binary search and linear search algorithms, not to compare and contrast inserting items into unsorted and sorted lists. 

Of  extraneous sentences in the student’s answer in Figure~\ref{fig:output_example1}, the sentences are categorized as 1 “Paraphrased Facts”. The student describes the difficulty of inserting items into an unsorted list in their own paraphrased words by answering \textit{“...and that makes it easier to add more values to the list”}. The student next describes the difficulty of inserting items into an sorted list in their own paraphrased words by answering \textit{“...which does make it harder to add new values”}. Both extraneous sentences in this student answer are in the “Paraphrased Facts” category. Therefore, the student receives 1 “Paraphrased Facts” for the presence of this kind of text in their answer.

In Figure~\ref{fig:extraneousness}, we see that participants using Aida included the  fewest extraneous sentences in their answers. The median extraneousness using Aida was 0.0 and mean was 0.17. The answers of students using ChatGPT had a median of 0.1 and mean of 0.19, while students using Google had an even higher extraneousness median of 0.2 and mean of 0.27. 
Table~\ref{tab:significance1}'s results of our ANOVA analysis ($p = 0.324, p > 0.1$) showed no significant difference between the means of the three conditions across both tasks in student answer extraneousness.
When we examined the type of extraneous sentences, 
we observed that using Google resulted in a very high relative proportion of copy/paste extraneous sentences.
In addition, Google student answers exhibited slightly fewer than the others' sentences that contained reflection and paraphrased facts.





\subsubsection{Effectiveness of Providing Examples to Students}
In addition to the answers that students provided as the outcome of their learning, we examined the effectiveness of each tool during the actual learning journey. 
Learning occurs through the presentation of real-life examples to the user~\cite{flax2023strategies, rawson2015power, denny2023computing}, so we examine the number of examples provided by the agent. Providing examples alongside the explanation of concepts bolsters and reinforces learning through demonstration of key characteristics associated with the concept. We show the number of examples provided by each tool per student session in Figure~\ref{fig:examples}. The results show that Aida is consistently showing at least one example to students in each session (median of 1, mean of 1.6). ChatGPT and Google, similar to each other, show examples slightly less often and less consistently. In addition, Aida presented examples in 73.33\% of total sessions while ChatGPT and Google presented in 43.48\% and 45.45\% of total sessions, respectively. These percentages represent specifically the proportion of sessions in which each tool provided relevant, illustrative examples in response to students' inquiries.
However, our ANOVA test ($p = 0.604, p > 0.1$) in Table~\ref{tab:significance1} showed no significant difference between the means of the three conditions across both tasks in presented examples to students.

\subsubsection{Effectiveness of Customizing to the CSP Course Scope}

As a proxy for the effectiveness of customizing the user's exploration to the CSP course scope, we further analyzed the student’s learning journey by measuring the appropriateness of the agent’s responses in relation to information intentionally excluded from the CSP course curriculum.


\begin{figure}[t]
\centering
\subcaptionbox{Number of Presented Examples\label{fig:examples}}{\includegraphics[width=0.45\textwidth]{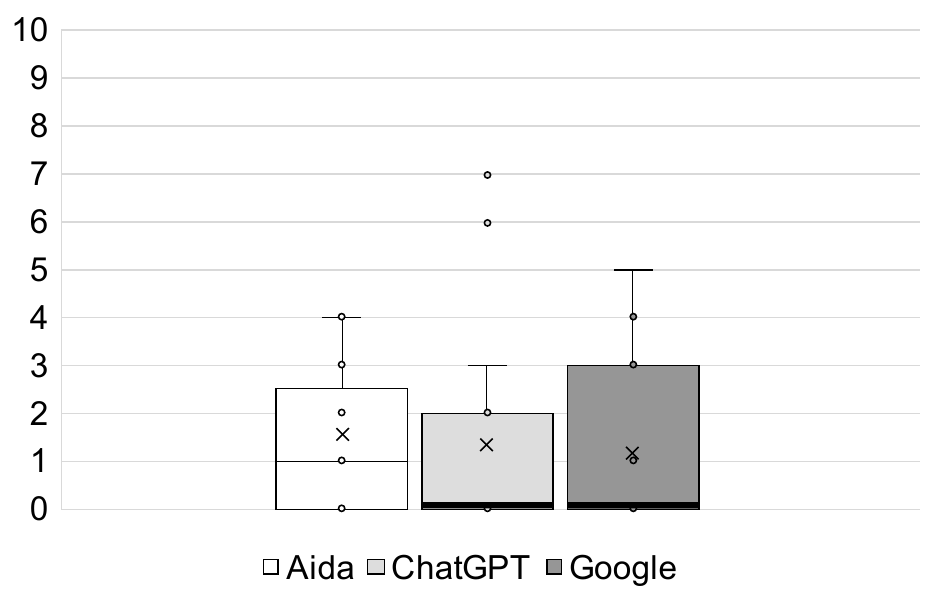}}%
\hfill
\subcaptionbox{Amount of Out of Scope Information\label{fig:out-of-scope}}{\includegraphics[width=0.45\textwidth]{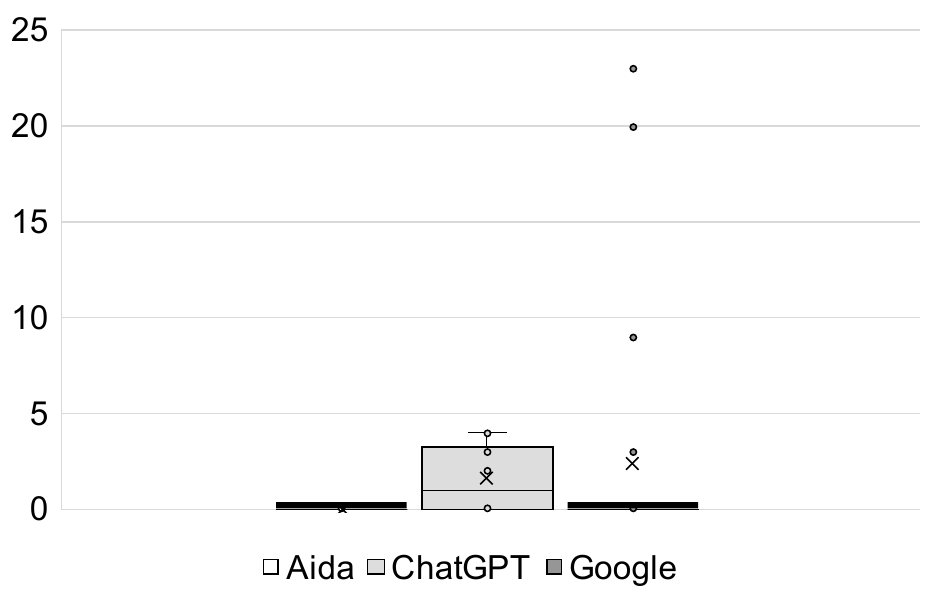}}%
\caption{Distribution of the Number of Presented Examples (left) and Computer Science Principles Exclusion Statement Occurrences per Session (right).  The results are aggregated across both tasks for each tool. All participants utilized Aida for one task, and either ChatGPT or Google for one task (Aida N=45, ChatGPT N=22, Google N=23).}
\label{fig:effectiveness_help_learn}
\end{figure}

The AP College Board CSP Framework defines exclusion content that does not need to be included in the course via specific ``exclusion statements''. The content described in the exclusion statements should not be covered by CSP and will not be assessed on the AP Computer Science Principles Exam~\cite{ap}, e.g., see Table~\ref{tab:exclusion-statements}. 
When exploring a task with a conversational agent, we considered the agent’s responses to be ``out of scope” if the agent reveals information that is within CSP exclusion statements. When exploring a task with web search, we considered the information retrieved from web sources to be ``out of scope” if the web page visited or search query contains information that is within the AP College Board’s exclusion statements for the task being explored. We counted the number of ``out of scope” information in each session of each tool in Figure~\ref{fig:out-of-scope}. It shows that ChatGPT usually generates the most sentences on topics in the exclusion statements per student session. While Google and Aida median number of ``out of scope'' instances is low, Google has a few student sessions with very high occurrence of exclusion statement content. We conducted an ANOVA comparing the students in the three conditions regarding the ``out of scope'' information presented during their sessions. Our ANOVA analysis (i.e., Table~\ref{tab:significance1}) indicates that students who interacted with Aida were exposed to significantly less ``out of scope'' information compared to ChatGPT ($M_{Aida} = 0, M_{ChatGPT} = 1.59, t = 4.297, Cohen's\:d_{s} = 1.612, p < 0.025$) having a large effect size. A subsequent comparison in our ANOVA analysis in Table~\ref{tab:significance1} shows that students were exposed to significantly more ``out of scope'' information with Google compared to ChatGPT ($M_{Google} = 2.39, M_{ChatGPT} = 1.59, t = 0.582, Cohen's\:d_{s} = 0.17, p < 0.025$) having a small effect size.

\begin{table}[h]
\caption{Advanced Placement College Board Computer Science Principles Exclusion Statements.}
\small
\begin{tabular}{|p{0.15\columnwidth}|p{0.8\columnwidth}|}
    \hline 
    Binary Search & (EK AAP-2.P.1) Specific implementations of the binary search are outside the scope of the course and the AP Exam. \\
    \hline
    Linear Search & (EK AAP-2.0.1) Traversing multiple lists at the same time using the same index for both (parallel traversals) is outside the scope of this course and the AP Exam. \\
    \hline
    \multirow{3}{0.15\columnwidth}{Algorithmic Efficiency} & (EK AAP-4.A.3) Formal analysis of algorithms (Big-O) and formal reasoning using mathematical formulas are outside the scope of this course and the AP Exam. \\    & (EK AAP-4.A.9) Specific heuristic solutions are outside the scope of this course and the AP Exam. \\
    \hline
\end{tabular}
\label{tab:exclusion-statements}
\end{table}

\begin{figure}
\centering
\subcaptionbox{Aida Conversation Excerpt\label{fig:conversation_aida}}{\includegraphics[width=0.45\textwidth]{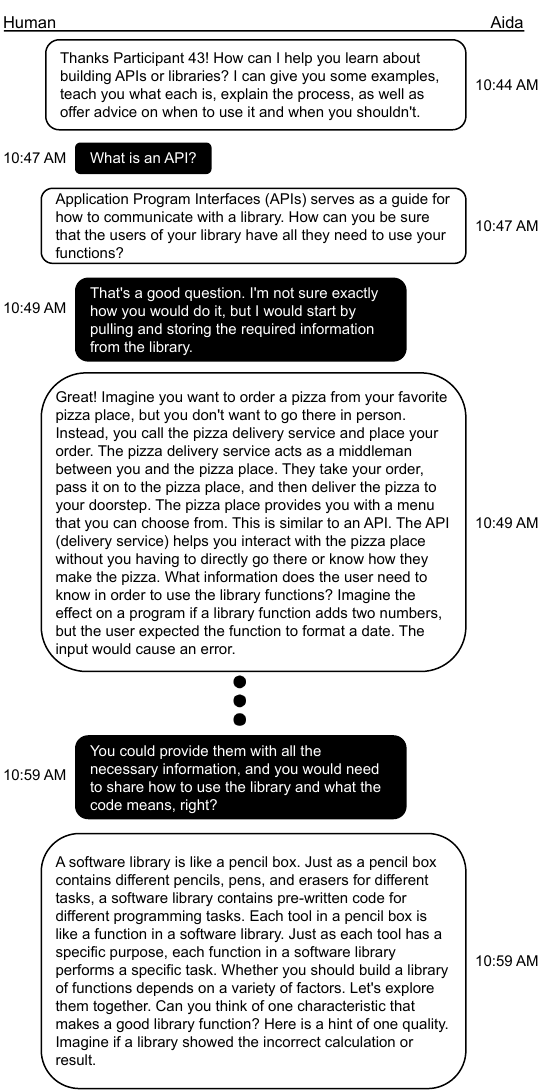}}%
\hfill
\subcaptionbox{ChatGPT Conversation Excerpt\label{fig:conversation_chatgpt}}{\includegraphics[width=0.44\textwidth]{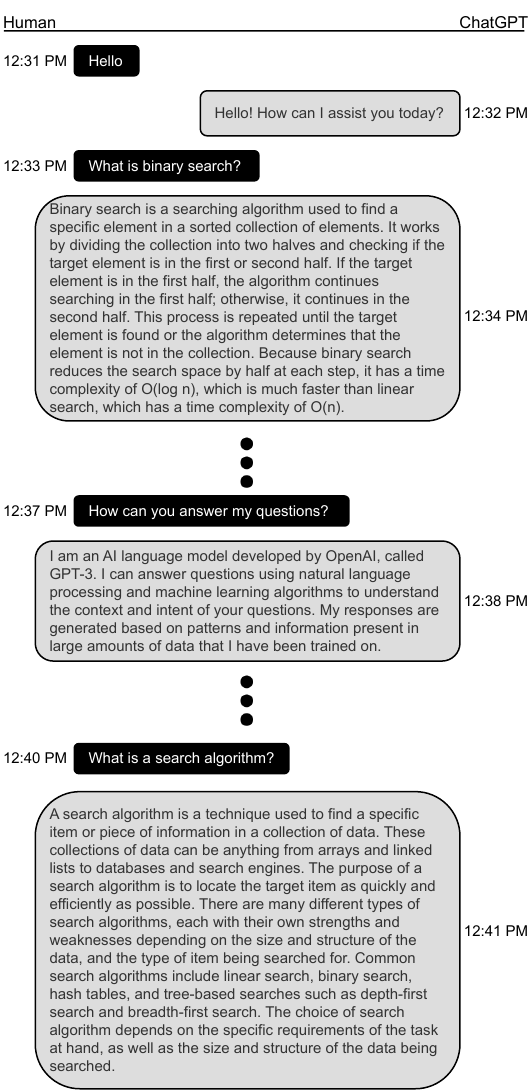}}%

\subcaptionbox{Google Search History Excerpt\label{fig:conversation_google}}{\includegraphics[width=0.95\textwidth]{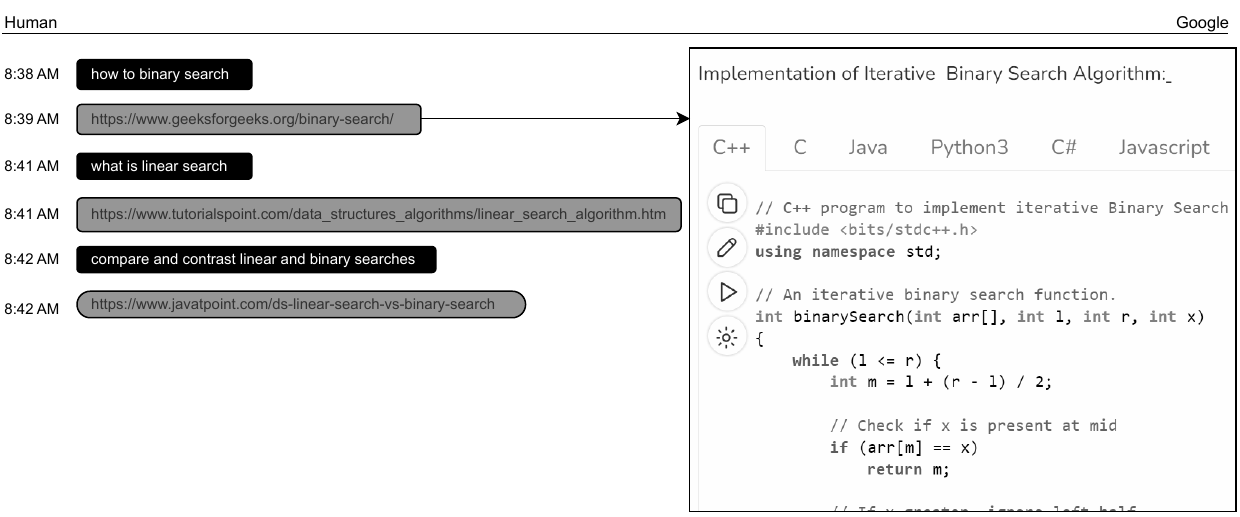}}%
\caption{Example student sessions:  Aida conversation,  ChatGPT conversation, and Google search history.}
\label{fig:conversation_examples}
\end{figure}

Figure~\ref{fig:conversation_examples} shows excerpts of student interactions with the tools. We designed Aida's fixed-responses to not include any ``out of scope” information. There are 3 ``out of scope” information in ChatGPT's responses during this student's exploration of the task. Not all interactions are shown, however, Figure~\ref{fig:conversation_chatgpt} shows 2 of the 3 ``out of scope” information included in ChatGPT's response. ChatGPT describes the binary search and linear search time complexity in Big O notation which violates the College Board Essential Knowledge Exclusion Statement EK AAP-4.A.3 as shown in Table~\ref{tab:exclusion-statements} when replying ``\textit{...Because binary search
reduces the search space by half at each step, it has a time
complexity of O(log n), which is much faster than linear
search, which has a time complexity of O(n).}''. Figure~\ref{fig:conversation_google} shows a student's Google search history including the search queries and websites visited when exploring the task. There are 20 ``out of scope” information on the websites visited during the student's exploration. While the websites that the student visited do include visual diagrams of both search algorithms, the websites also contain several coding implementations of binary search and linear search in iterative and recursive techniques in several programming languages, as well as Big O Notation for time complexity and space complexity for the search algorithms. These details are ``out of scope” and violate the College Board Essential Knowledge Exclusion Statements EK AAP-2.P.1 and EK AAP-4.A.3 as shown in Table~\ref{tab:exclusion-statements}.

\subsection{User Engagement}

To evaluate our second hypothesis (H2) that ``\textit{Generative conversational agents lead to better perceived student experiences despite being less engaging than fixed-response conversational agents}'', we analyzed the user engagement metrics alongside the participant post-study survey perspectives of the conversational agents. We also report a comparison to Google search for a comprehensive analysis to enhance our discussion in the subsequent section. We analyzed each student’s engagement with the tools, where engagement is defined as the extent to which a student is involved in the exploratory learning experience with the tool. We measured engagement through two metrics computed from their conversation logs. 
First, we analyze the number of their exploratory actions with each tool. Second, we assess their time spent on each task.

We define the number of student exploratory actions as the sum of all attempts that a student takes to continue achieving the task being explored. In sessions where the task is explored with a conversational agent, we consider the student’s “exploratory action” as the total number of messages that the student sends the conversational agent related to the task being explored. In sessions where the task is being explored with a conventional web search engine, we consider the student’s “exploratory action” as the total number of websites visited during their session (counting both search result and web content pages). Each student’s session received a score representing the count of the total number of exploratory actions for each task being explored.

We define the task duration as the amount of time that a student interacts with one of the tools to perform one of the exploratory search tasks. We measure the task duration as the difference between their last recorded interaction time and their first interaction time with a tool. In sessions where a task is being performed with a conversational agent, each speaker’s message receives a timestamp at the time that the message is sent. In sessions where a task is being performed with Google, the web browser’s history records the time when a web page is visited. Each student’s task duration was represented as the total number of minutes the student explored for a specific task.

For instance, while not all the interactions are shown for space, there are 9 student interactions related to exploring the task with Aida as shown in Figure~\ref{fig:conversation_aida}. The student either asks a question related to the task being explored (e.g., \textit{“What is an API?”}) or responds to Aida’s question (e.g.,\textit{ “That's a good question. I'm not sure exactly how you would do it, but I would start by pulling and storing the required information from the library.”}). 
The session duration for this conversation is 15 minutes, the difference between the last recorded interaction time in the conversation, 10:59 AM, and the first recorded interaction time, 10:44 AM. 
Figure~\ref{fig:conversation_chatgpt} shows excerpts of student interactions with ChatGPT. There are 4 student interactions with ChatGPT related to the task, i.e., student interactions such as \textit{``What is binary search?''} and \textit{``What is a search algorithm?''} are related to completing the task and counted as exploratory actions. Conversely, student interactions such as \textit{``Hello''} and \textit{``How can you answer my questions''} are unrelated to the task being explored and are not counted towards this student's exploratory actions. The session duration for this conversation is 10 minutes (12:31 PM - 12:41 PM). 
%
There are 6 student interactions with Google shown in Figure~\ref{fig:conversation_google}. The student either searches Google for the task being explored (e.g., \textit{“how to binary search”}) or visits a web page for information on the task being explored (e.g.,\textit{ “https://www.geeksforgeeks.org/binary-search/''}).
The session duration is 4 minutes (8:38 AM - 8:42 AM).

\begin{figure}[t]
\centering
\subcaptionbox{Number of Student Exploratory Actions}{\includegraphics[width=0.45\textwidth]{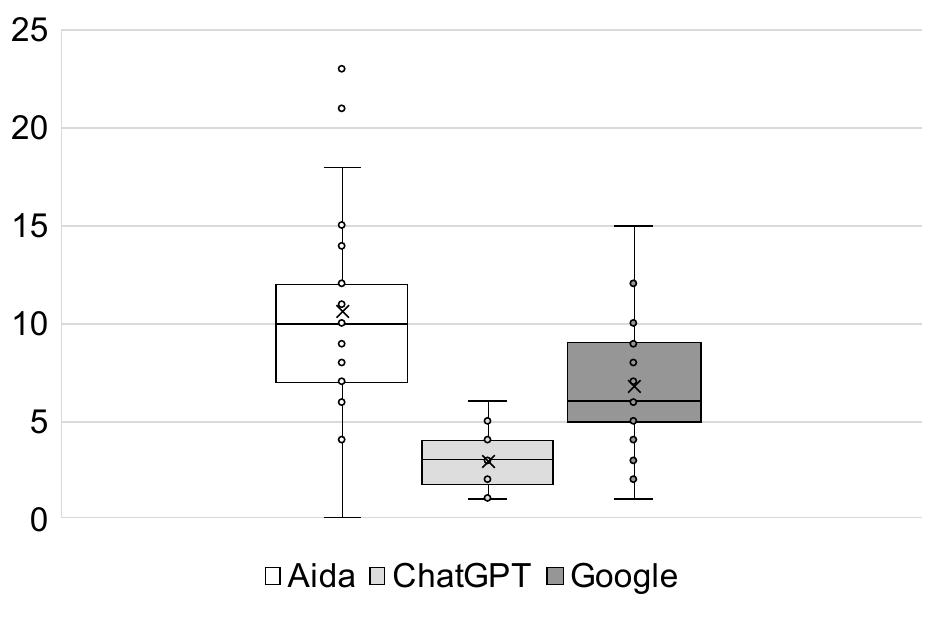}}%
\hfill
\subcaptionbox{Task Duration}{\includegraphics[width=0.45\textwidth]{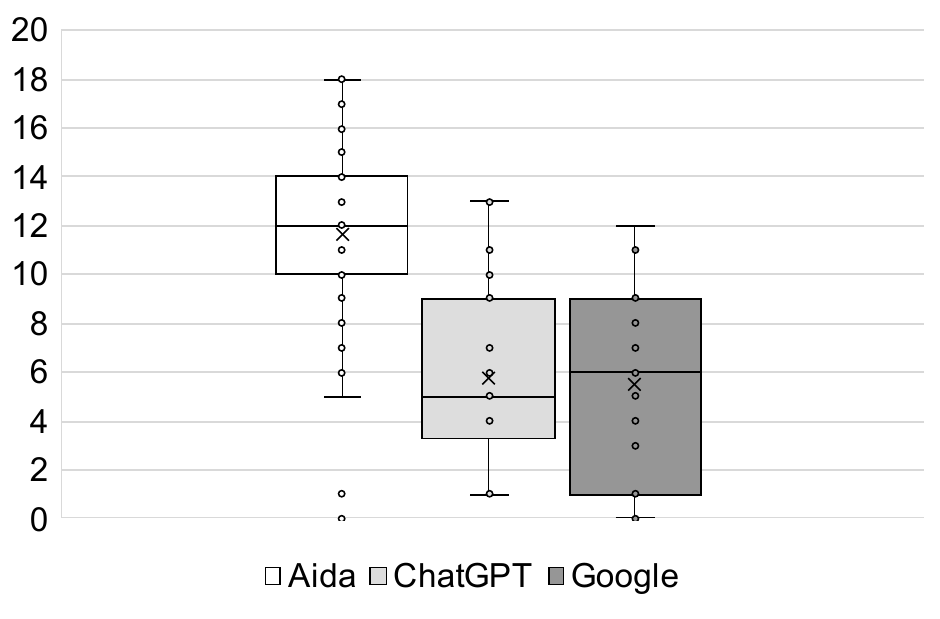}}%
\caption{Results regarding measures of user engagement including the Number of Student Exploratory Actions (left) and Task Duration (right). The results are aggregated across both tasks for each tool. All participants utilized Aida for one task, and either ChatGPT or Google for one task (Aida N=45, ChatGPT N=22, Google N=23).}
\label{fig:user_engagement}
\vspace{-3mm}
\end{figure}

Figure~\ref{fig:user_engagement} presents the results for the number of student exploratory actions and task duration per session over both tasks for all students. We see that Aida's engagement was higher than ChatGPT and Google. Students interacted with Aida more (median of 10 exploratory actions) compared to ChatGPT and Google (medians of 4 and 6 exploratory actions, respectively) and spent more time using the tool (median of 12 minutes). On the other hand, ChatGPT generated the fewest number of actions, by a wide margin, and the shortest task duration. As indicated in Table~\ref{tab:significance1}, results of our ANOVA analysis comparing the students in the three conditions regarding the task duration and number of student exploratory actions shows that students who interacted with Google completed their tasks significantly in less time than compared to ChatGPT ($M_{Google} = 5.478, M_{ChatGPT} = 5.773, t = -0.254, Cohen's\:d_{s} = -0.076, p < 0.001$) having a small effect size. Regarding the number of student exploratory actions, we found that students interactions with Aida produced significantly more exploratory actions compared to Google ($M_{Aida} = 10.556, M_{Google} = 6.783, t = 3.663, Cohen's\:d_{s} = 0.827, p < 0.001$) and student interactions with Google produced significantly more exploratory actions compared to ChatGPT ($M_{Google} = 6.783, M_{ChatGPT} = 2.909, t = 5.012, Cohen's\:d_{s} = 1.472, p < 0.001$), both having a large effect sizes (i.e., Table~\ref{tab:significance1}).



\subsection{Student Perspectives}

After all students completed both tasks, each student answered a survey to obtain their perceptions of the tools in terms of effectiveness and the interactiveness of the two kinds of conversational agents. Each participant used only two of the exploratory search methods.
We conducted an Wilcoxon Rank-Sum Test comparing the students' ordinal Likert scale responses of the two conversational agents across both tasks for all post-survey metrics of effectiveness and interactiveness as shown in Table~\ref{tab:significance2}.

\begin{table}[h]
\small
\caption{Overview of results (mean and standard deviation (SD)) of post-study survey on conversational agent preference regarding effectiveness (i.e., providing useful information, helpfulness, using it in the future) and engagement (i.e., understandability, understanding the user, behaving awkwardly, number of questions asked) across both tasks.}
\begin{tabular}
{||p{0.1\columnwidth}|p{0.015\columnwidth}|p{0.09\columnwidth}|p{0.09\columnwidth}|p{0.09\columnwidth}|p{0.09\columnwidth}|p{0.1\columnwidth}|p{0.1\columnwidth}|p{0.09\columnwidth}||}
\hline 
\multicolumn{2}{||c}{} & \multicolumn{3}{|c}{Effectiveness} & \multicolumn{4}{|c||}{Interactiveness} \\
\hline 
\multirow[t]{4}{0.015\columnwidth}{\bf Tool} & \multirow[t]{3}{0.02\columnwidth}{\bf N} & \textbf{Useful Information} & \textbf{Helped Me} & \textbf{Future Use} & \textbf{Understand Tool} & \textbf{Understand Me} & \textbf{Behaved Awkwardly} & \textbf{Too Many Questions} \\
& & mean & mean & mean & mean & mean & mean & mean \\
& & (SD) & (SD) & (SD) & (SD) & (SD) & (SD) & (SD) \\
\hline
\multirow[t]{2}{0.015\columnwidth}{Aida } & \multirow[t]{3}{0.02\columnwidth}{45} & 3.53 & 3.31 & 2.51 & 3.36 & 2.71 & 3.09 & 3.58 \\
 &  & (.87) & (1.08) & (1.16) & (1.03) & (1.06) & (1.24) & (1.20) \\
 \hline
 \multirow[t]{2}{0.015\columnwidth}{ChatGPT} & \multirow[t]{3}{0.02\columnwidth}{22} & \textbf{4.41} & \textbf{4.59} & \textbf{4.50} & \textbf{4.32} & \textbf{4.36} & \textbf{2.45} & \textbf{1.32} \\
 &  & \textbf{(0.8)} & \textbf{(0.59)} & \textbf{(0.6)} & \textbf{(0.78)} & \textbf{(0.66)} & \textbf{(0.91)} & \textbf{(0.57)} \\
\hline
\multirow[t]{3}{0.015\columnwidth}{\textbf{Significance}} & \multirow[t]{3}{0.02\columnwidth}{} & p \textless 0.005 & p \textless 0.001 & p \textless 0.001 & p \textless 0.01 & p \textless 0.001 & ns & p \textless 0.001 \\
(Cohen's ds, t-value) & & (-1.35, -4.10) & (-1.34, -6.25) & (-1.96, -9.26)  & (-1.01, -4.26)  & (-1.74, -7.83)  &   & (2.18, 10.48)  \\
\hline
\end{tabular}
\label{tab:significance2}
\end{table}

\subsubsection{Quantitative Insights. }
\begin{figure}
    \includegraphics[width=0.95\linewidth]{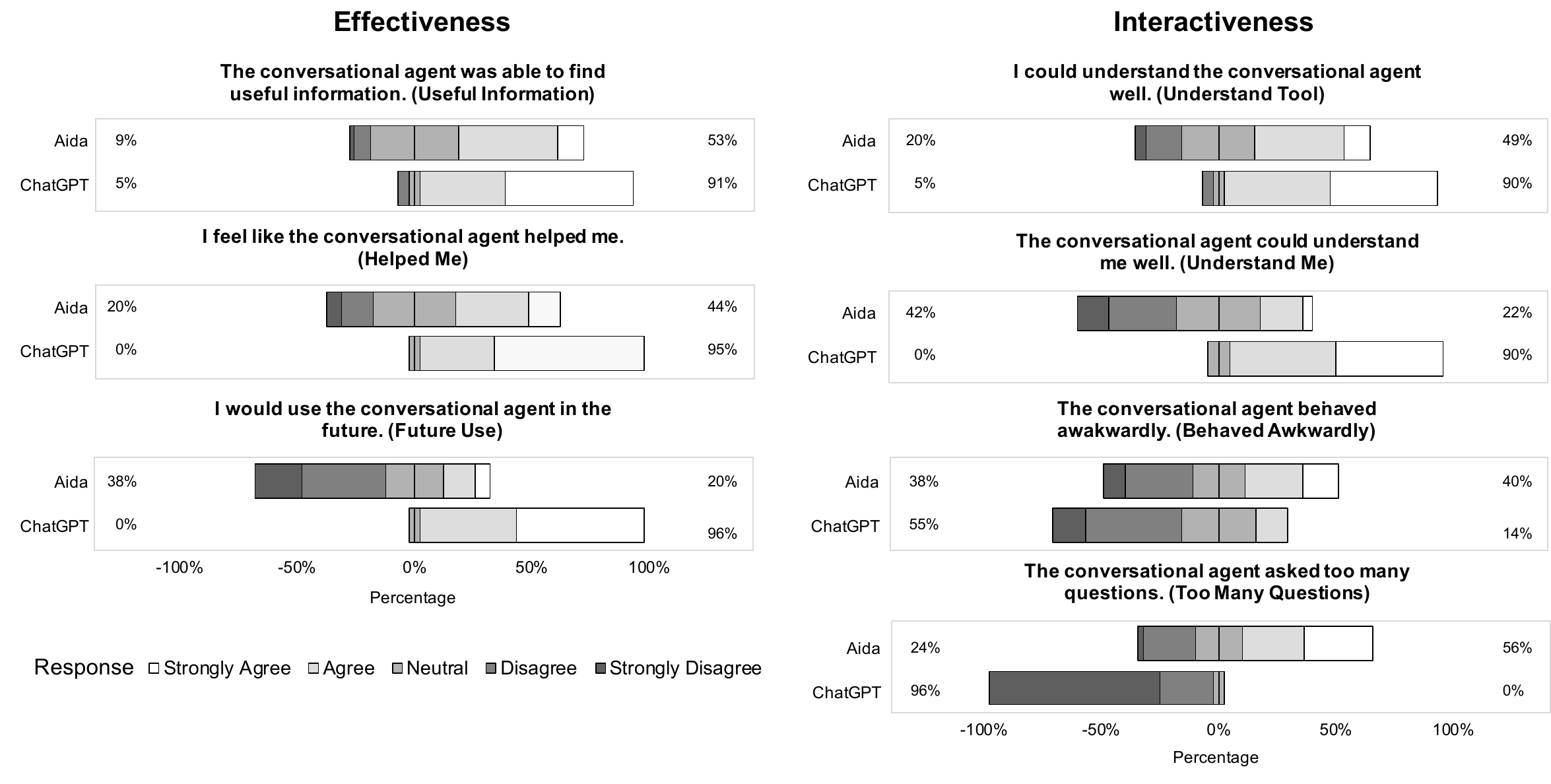}
	\caption{Students' Perception of Conversational Agent Interactions.  All participants utilized Aida for one task, and either ChatGPT or Google for one task (Aida N=45, ChatGPT N=22, Google N=23).}
	\label{fig:likert_combined}
\vspace{-3mm}
\end{figure}

Figure~\ref{fig:likert_combined} presents the results of survey questions about the two conversational agents' (Aida and ChatGPT) interactions and effectiveness. Student perspectives on ChatGPT are considerably more favorable than Aida for all of these questions: understanding the user, understandability, behaving awkwardly, number of questions asked, providing useful information, helpfulness, and using it in the future. There is a clear participant preference towards ChatGPT based on these results. There was a significant difference among students for each survey question with regard to preference of ChatGPT over Aida ($p < 0.01$) as shown in Table~\ref{tab:significance2} with the exception of student's perceptions of the conversational agent that behaved awkwardly, each having large effect sizes. There was no significant difference between the conversational agent in terms of displaying awkward or non-humanlike behavior ($p > 0.2$).

The survey results from questions comparing the tools are shown in Table~\ref{tab:user_experience}.
It is notable that for each question, each tool was preferred by at least some subset of the participants. ChatGPT was chosen by the highest percentage of participants as enabling them to understand more based on their personal experience, making it easier to complete the task, learning more from the tool, and providing more enjoyment than Aida or Google Search.  Aida, the fixed-response conversational agent was chosen the lowest percentage of the time for all these characteristics.

\begin{table}[h]
\small
\caption{Overview of students' comparisons on their preferences of tools regarding the method that enabled understanding more based on personal experience, the method that made it easier to complete the task, the method that they learned more from, and the method that was more enjoyable to use. All participants utilized Aida for one task, and either ChatGPT or Google for one task (Aida N=45, ChatGPT N=22, Google N=23).}
\begin{tabular}
{||l|c|c|c|c||}
\hline  
{\bf Tool} & \textbf{Personal Experience} & \textbf{Easiest To Use} & \textbf{Learn More} & \textbf{Enjoy More} \\
\hline
Aida & 20\% & 16\% & 16\% & 13\% \\
\hline
ChatGPT & \textbf{29\%} & \textbf{44\% }& \textbf{40\%} & \textbf{42\%} \\
\hline
Google & 24\% & 36\% & 31\% & 29\% \\
\hline
No Preference & 27\% & 4\% & 13\% & 16\% \\
\hline
\end{tabular}
\label{tab:user_experience}
\end{table}

\subsubsection{Thematic Analysis}
The researcher of this paper used a reflexive process of thematic analysis, as described by Braun and Clarke~\cite{braun_2021_thematic}, to analyze the open-ended question responses in our post-study survey. The reflexive analysis process is a theoretically flexible method to developing, analyzing, and interpreting patterns in qualitative data. 
Reflexivity involves drawing upon your experiences, pre-existing knowledge, and social position and critically interrogating how these aspects influence and contribute to the research process and potential insights into the data~\cite{devine_2021_reflexive}. 
Given that the qualitative data set in our study was three open-ended questions per student for each tool (i.e., Aida, ChatGPT, Google), the researcher labeled parts of the student responses that captured important features of the data. The process involved organizing the open-ended responses by question and systemically coding each response by question. Lastly, the labels were grouped into broader themes for reporting and discussion.

\paragraph*{Qualitative Insights. }
\ We asked students to describe what they liked and disliked about their experience (i.e., effectiveness and interactivity) completing the task, as well as to provide any other additional comments on each tool. 
The students' open-ended responses indicate their positive perception of the effectiveness of conversational agents, and do not support positive perception of the effectiveness of conventional web search. Here, Px refers to Participant x. 

Participants responded positively to Aida's responses and particularly enjoyed the examples that Aida provides.  
P60 
writes ``\textit{It[Aida] felt like a conversation and it was smooth, it gave me answers and connected the answers to real world things that I could relate to which was helpful.}'').
Students also expressed positive perceptions of ChatGPT responses' effectiveness. 
P79 writes ``\textit{I liked how ChatGPT would give me what I needed while give even more info which let me ask more and better questions}''. 
The most common response from students related to Google's effectiveness was that the information retrieved from Google was too difficult to understand. For instance, P90 writes 
``\textit{There was no easy way to condense the information into small, readable chunks. Also, there was no way to get the websites to explain the concept in a way I might understand better, just the one explanation they have for everyone.}'' 

Further evidence in the open-ended responses shows that students were skeptical of the credibility of websites from their Google search (i.e., P64 writes ``\textit{One thing I think I disliked was that, since there were so many sources all right next to each other, it was sometimes troublesome to decide if one was more correct/legitimate than the other.}''). 
However, compared to the credibility of the information retrieved from web search, students' perceptions were that the information retrieved from both conversational agents was credible.
P78 writes ``\textit{I liked that Aida was specifically for programming and seemed to only have information about what I was searching for. It made it so everything was concise and I knew that whatever I was searching for would have an answer.}'' No students commented on not trusting the credibility of ChatGPT with the exception of P43 who observed that ChatGPT hallucinated.

The students' open-ended responses support the perception of the interactions with conversational agents and conventional web search provided in the closed question results presented earlier. The responses about Aida's interactions were mixed and polarized, where the primary theme revolved around 
the pedagogical design of Aida compared to ChatGPT, as well as 
Aida's intent matching. 
Specifically, some students really enjoyed that Aida guided them to arriving at their own answers without divulging the answer while others did not enjoy the frequency of Aida's questions. 
P75 dislikes Aida's interactions writing ``\textit{I disliked how linear it[Aida] felt, like it was searching for an exact answer and if it didn't receive it, it would just throw the question back to me with a limited amount of change. It made getting information hard, as I would answer to the best of my ability and be roadblocked. When I tried to ask it other questions related to the subject, it again railroaded me onto the track of questions I had first received. In conversation, a real human wouldn't just keep saying the exact same thing back to someone when they are trying to answer questions.}''.
Responses regarding the infrequency of Aida matching the appropriate DialogFlow intents include
P66 who writes ``\textit{I feel like Aida didn't always understand what I was saying and there was some confusion}''. 
Nearly all students interactions were positive with ChatGPT. 
While students did largely enjoy the responses from ChatGPT, two students commented that ChatGPT required additional prompt engineering to appropriately gather the information that they were seeking. For instance, P89 writes ``\textit{i had to be very specific to find what i was looking for}''.

Those students who enjoyed the interaction with Aida commented vastly different. 
P58 writes ``\textit{I think that this[Aida] is a great educational tool, and that it is almost like an online teacher. The answers it gave were also easy to understand and comprehend.}'' 
P78 enjoyed Aida's Socractic questioning writing ``\textit{I also like that it[Aida] was friendly and would sometimes ask questions that test your understanding and make you think. It made it so I could contribute and made me comfortable in knowing that it was providing helpful, relevant information.}''.
P64 writes ``\textit{I liked that fact that, unlike when I was using the Google search browser, AIDA was more interactive than of course the search browser did. Being able to drive a discussion about what I was researching made things easier in some cases.}''. 
Nearly all students enjoyed Aida's speed of responses (i.e., fixed-response intent matching) 
compared to the speed of ChatGPT's generative responses.
Regarding ChatGPT, 
P69 writes ``\textit{It[ChatGPT] took up to 30 seconds to respond}''. 
On the other hand, regarding Aida, P59 writes ``\textit{The responses were very quick and almost instantaneous}''. 


Not surprisingly, students' feedback supports their positive perception of the interactions with conventional web search, yet interestingly students voiced concerns with the struggles of information literacy in web search. Many students acknowledge that they are already familiar with using Google and, therefore, they found it easy to use (i.e., 
P54 writes ``\textit{I liked using google because I am more familiar with how to ask an efficient question.}''). 
While nearly all students enjoyed the breadth of sources that conventional web search provides, the variety of sources was a double-edged sword that made it difficult to find useful information and the search process overwhelming for students (i.e., 
P86 writes ``\textit{The amount of sources for the task were few and far between.}'').
P70 expressed a considerable amount of query reformulation in their Google queries by writing ``\textit{I disliked that it was sometimes difficult to get Google to show me the answer I was looking for. Sometimes I had to try to reword my question multiple times to get the information I was looking for.}''. 

\section{Discussion}


This study aimed to investigate two research
hypotheses: (H1) Conversational agents provide a better learning experiences than conventional web search for secondary school CSP students performing exploratory search, and (H2) Generative conversational agents lead to better perceived student experiences despite being less engaging than fixed-response conversational agents. The effectiveness and engagement results as well as qualitative analysis of our study support both hypotheses.
Our study reveals a few salient points related to the use of conversational agents in CSP, as the computing education field copes with the rapidly evolving technology that generative AI provides.
We present a discussion of opportunities and challenges in leveraging the benefits of conversational agents while mitigating risks.

%



\subsection{Opportunities In Using Conversational Agents For CSP Exploratory Search}
First, conversational agents have the opportunity to act as a single knowledge source (i.e., personalized tutor) for students who struggle with their conventional information gathering experience. We observed that ChatGPT performed well on the effectiveness measures, i.e., it produced the most complete answers, and was by far the most liked tool by the students. Clearly, this supports our first hypothesis (H1) and indicates that LLMs have a lot of potential for use in student exploratory search for information related to CSP. However, ChatGPT was outperformed by Aida in terms of the higher accuracy and lower extraneousness of the answers. This indicates that Aida's custom responses lead to high quality of learned information by the students. Nonetheless, Aida's rigid conversational structure was generally disliked by the students and supports our second hypothesis (H2). Relative to both of the conversational agents, Google performed poorly across the board in effectiveness as well as in student perceptions. The learning tasks in our study were intentionally designed to span across multiple searches and require exercising information literacy skills such as locating, aggregating, organizing, and communicating acquired knowledge. In their experiences with Google search, students looked for a single web source that consolidated the information to complete the task. 

Second, conversational agents present an opportunity to leverage the existing trust that students have in them and provide an enhanced learning experience for students.
Students exhibited high levels of trust in conversational agents compared to information retrieved from conventional web search. Unlike Aida, there is a potential for ChatGPT to hallucinate. Although hallucination is far from a solved problem, given how powerful AI models are becoming and the rapid pace at which new models are being released, we anticipate that the amount of hallucination will become less of a problem in the near future. Only one student questioned the credibility of ChatGPT. Comparatively, students questioned the information retrieved from websites in their Google search exploration. Beyond our study, CS educators are increasingly worried that their students will become over-reliant on LLMs for automated responses~\cite{prather2023robots}. Notably, although this was not the case that we observed in our study, CS educators fear ``blind trust'' where students do not question LLMs responses given how advanced their natural language capabilities are.

Additionally, the discrepancy in students' training to evaluate information from web search results versus conversational agents like ChatGPT is a significant factor in understanding their trust in these tools. While students are often taught to critically assess the credibility of web search results, similar instruction for conversational agents is generally lacking. This educational gap may contribute to the higher level of trust observed in responses generated by ChatGPT, despite the potential for inaccuracies. As conversational agents become more integrated into educational settings, it is crucial to provide students with explicit instruction on how to critically evaluate the information produced by these tools, just as they would with traditional web search results. Addressing this gap will better equip students to navigate the complexities of information retrieval across different platforms and foster more informed use of AI-driven technologies.

Third, conversational agents have the opportunity to broaden participation in CS. Students reported that all exploratory learning methods enabled them to learn based on their personal experience, relatively equally. This indicates that students of all backgrounds can utilize learning with conversational agents.
Further, there is an opportunity for conversational agents to reduce the workload of instructors by providing personalized help to students who are struggling and who would otherwise consume a considerable amount of instructor time~\cite{prather2023robots}. Given a more personalized experience, the inherent biases of LLMs should be considered when broadening participation (e.g., the impact on the model's behavior to differing ethnic student names when added to a prompt)~\cite{prather2023robots}. 

We also observed that ChatGPT did not ask any clarifying questions to students during any of the sessions. However, ChatGPT has the ability to ask clarifying questions, which could be an opportunity for the agent to gather information about its audience in an effort to better tailor its responses.  The lack of clarifying questions is more akin to a conventional information retrieval or recommendation system as opposed to a conversational agent, which contradicts its name, ``ChatGPT''. Incorporating timely and effective Socratic questioning into ChatGPT through clarifying questions could better align LLMs to the CSP pedagogical framework. In contrast, Aida asked clarifying questions based on the defined intents. 

\subsection{Challenges In Using Conversational Agents For CSP Exploratory Search}
Despite ChatGPT's high effectiveness for task completion, using LLMs for exploratory search poses several challenges as a learning tool. The first challenge is that LLMs do not scope their responses to their audience because they are a general-purpose tool for broad audiences. Unbeknownst to the students in our study, ChatGPT provided the highest count of sessions with ``out of scope'' responses (most consistent), which were deemed inappropriate to include in CSP according to the College Board's Curriculum Framework, relative to the other tools. Google provided the largest average amount of ``out of scope'' responses, however, the mean was driven by outlier sessions with high quantities of ``out of scope'' responses. Further, students reported that when using Google search, they largely did not understand the information retrieved which is indicative of ``out of scope'' knowledge. Without guardrails or additional customization, the information retrieved from generative conversational agents have the potential to become too difficult for secondary school students to understand as well.
To this end, in the same manner that students should learn how to effectively use Google search, students using LLMs for exploratory search should be instructed in prompt engineering, the process of effectively structuring text so that it can be interpreted by an AI model. 
Comparatively, at the expense of being less human-like, Aida structures the dialogue to explore topics sequentially. 

Another challenge relates to convenience in completing the task.  Students favor convenience in their learning journey and largely agree that ChatGPT is the easiest and shortest duration method to complete the task. It can be perceived that speed and ease of use are benefits of LLMs. However, CS educators believe LLMs pose a risk of creating ``lazy learners'' in introductory CS educational settings (i.e., Do students actually learn more using generative conversational agents or do students agree that their learning is ``better'' because it is easier?)~\cite{prather2023robots}. There are advantages to a high degree of natural language processing and natural language understanding such as decreased task duration. However, there are also tradeoffs, such as a decreased number of exploratory actions. 
Expert interviews indicate that CS educators believe students should learn the core concepts in introductory CS courses prior to expanding learning to leverage generative AI in higher level CS courses~\cite{prather2023robots}. Especially considering that students may be required to use LLMs in professional settings, yet LLMs will not be available during job interviews, students should have some level of exposure in higher level course settings to prepare them for a successful transition into the workforce. Students in our study agree that utilizing Google to complete the task was easy; however, their opinions were influenced by having prior familiarity with already using the tool. Students noted that Aida was the most challenging to use (i.e., the least convenient). Interestingly, the Think-Pair-Share pedagogical approach is already implemented within in the CSP course lectures and activities, however, the students did not prefer it when it came to their own individual information gathering. With more exposure to gaining knowledge using the Think-Pair-Share pattern through conversational agents, there is a possibility students will adapt and change perceptions in future iterations of the tool whilst mitigating the challenge of creating ``lazy learners''. 

The next challenge is the ethical implications of utilizing LLMs in an educational context. In our study, we observed that in some cases students entered the full task prompt or large portions of the task prompt directly into both conversational agents. This level of effectiveness has risks hindering the efficacy of using LLMs as a learning tool and imposes ethical concerns on the outcomes of student's tasks. 
This is evidenced by students presenting a model's answers directly as their own. Nothing prohibits LLMs from completing the task directly for the student as opposed to supporting the student to explore completing the task. Research shows that presenting generative AI's knowledge as your own ideas is not plagiarism, yet it constitutes a form of falsification of the student's ideas~\cite{prather2023robots}. Further, many schools have not yet produced well-formed academic policies on LLMs. As a result, schools simply prohibit use of LLMs for learning, which poses risk of students violating unauthorized resource usage policies since CS educators believe that students will use LLMs anyway~\cite{prather2023robots}. While in both Aida and Google, we also observed that students similarly directly reported parts of the information retrieved as their own. However, there were no intents in Aida to match entering the full task prompt in order to generate a full succinct answer. Similarly, only a few students responded that they were able to eventually navigate to a single website that was able to provide enough information they needed to complete task; the majority of students were unsuccessful at finding a single website with a full succinct answer. 
\section{Threats to Validity}

Results may vary with different tools representing fixed-response and generative conversational agents and search engines.
In particular, the high number of ``out of scope'' responses from the generative agent could correspond to the choice of model. At the time of the study (May 2023), GPT-4~\cite{openai2023gpt4}, whose training data is significantly greater than GPT-3.5, was not yet publicly available. We mitigated this risk by using the largest GPT conversational model publicly available at the time of the study. Further, the lack of clarifying questions can be fine-tuned through prompt engineering. We mitigate this risk by allowing students to provide any input into the model.
A fixed-response dialog structure inherently imposes a constraint to multi-turn exploratory conversations, which will, by design, appear less engaging and less human-like compared to generative conversational agents. 
Due to the potentially wide variety of intents that the intended users may have, there is always a possibility a user may ask a question that is not supported by the fixed-response conversational agent. To mitigate this, each Aida model was trained on a variety of diverse phrases for each intention, and each intent contained a variety of fixed responses to reduce the likelihood that a student would receive the same response. We also piloted Aida's interactions, the study procedure, and survey on several non-authors and made improvements. 

Recognizing the spectrum of AI design possibilities, we focused on the specific configurations of Aida and ChatGPT that reflect deliberate design choices by OpenAI for ChatGPT and the pedagogy used by Code.org for Aida. Although alternative designs could have led to different outcomes, we also found the default configurations compelling based on our own experimentation. Future work could explore varying system prompts, customization options, and student interaction patterns to further understand the impact of AI design choices on learning experiences.

We acknowledge that the number of examples presented to a learner is only a proxy for the effectiveness of helping users learn. We cannot guarantee that the user read or understood the examples presented. Yet, Denny et al.~\cite{denny2023computing} proposes that exemplar solutions from generative AI in computing education has the potential opportunity to benefit learners by saving time from instructors on generating examples~\cite{3569823} and helping learners appreciate the trade-offs of different solutions approaches~\cite{3387425}. 

To mitigate against threats to construct validity, we constructed our exploratory tasks based on Athukorala et al. which lend themselves to exploratory search~\cite{10.1002/asi.23617}. The tasks were chosen because they are more complex topics in the CSP curriculum that motivate exploratory learning.
Internal threats due to possible task/tool use ordering bias were minimized by assigning the same tasks to different participants in different orders and changing the tool use order. 
Interpretation bias in the qualitative analysis of Student Answer Completeness Criteria was minimized by multiple experts creating the criteria, yet future studies could involve multiple coders or a consensus-based approach to enhance reliability of the open-ended survey.
The main external threat is generalizability beyond our participant set.  We recruited students from various CSP sections in three high school CSP courses across grades 9-11 (ages 14-17). Replication to broader demographics is encouraged. We focused on exploratory search where more query reformulation is common; our results may not generalize to other kinds of search.

Students may have had previously-held, strong opinions of the exploratory tools Google and ChatGPT. Google is a well-known web search engine that students likely have experience using. ChatGPT is, at the time of writing, one of the most highly publicized and widely recognized AI models available. Its significant media coverage, coupled with the broad popularity of Google’s products, may have influenced and potentially skewed students’ perceptions of its effectiveness in assisting with CSP tasks. This increased visibility and prominence of ChatGPT and Google could lead to an inherent bias in evaluating its utility relative to other tools, such as Aida. It is crucial to account for this potential bias, as the overwhelming visibility of ChatGPT and Google might unduly affect judgments regarding the comparative efficacy of various AI tools. 
Lastly, we acknowledge that the timing of this field study in May 2023, when most CSP topics had likely already been covered, presents a threat to the study's validity. This may have influenced the students’ interactions with the AI tools, as their prior knowledge could have affected their engagement and the outcomes observed.

The findings of this study, while specific to Computer Science Principles (CSP), could have broader implications for other disciplines where generative AI-driven educational tools are utilized. Challenges related to student engagement, response accuracy, and the need for prompt engineering are relevant across fields such as mathematics, biology, and the humanities, where similar technologies are increasingly being explored. Furthermore, the tools selected for this study serve as representatives of broader categories of AI applications, suggesting that the insights gained may inform the design and implementation of similar systems in other educational contexts.

\section{Conclusions and Future Work}
Our study aimed to investigate the potential of conversational agents in aiding high school students as they acquire knowledge on CSP concepts through exploratory search. We found that conversational agents overall have the potential to help students during the learning process. Specifically, students prefer conversational agents compared to conventional web search. The shorter task duration, less exploratory actions, and higher average extraneous information coupled with the proportionally large amount of copy/paste information using  web search promotes the use of conversational agents that are more adept to a pedagogically sound framework for learning CSP at the secondary school level. Furthermore, our results indicate that generative conversational agents are highly effective and interactive, and are preferred over fixed-response conversational agents, yet generative conversational agents pose risks as learning tools. Students are dissatisfied with the interaction and effectiveness of fixed-response conversational agents; however, find them helpful at finding useful information. 

Based on our findings, future work can focus on bringing the benefits of the fixed-response conversational agent and generative conversational agents together. Specifically, we plan to investigate approaches to customize a hybrid-domain conversational agent for learning CSP concepts.
Additionally, we acknowledge that ecological validity and student engagement are critical factors in evaluating the effectiveness of AI-driven learning experiences. Although this study did not include a longitudinal evaluation, we propose it as a future research direction to better understand the long-term effects of such tools on student learning.
Our study acknowledges a spectrum of AI design possibilities, and while we focused on specific configurations of generative and fixed AI systems, we recognize that alternative designs could have led to different outcomes. Future work could explore varying system prompts, customization options, and student interaction patterns to further understand the impact of AI design choices on learning experiences.


\bibliographystyle{ACM-Reference-Format}
\bibliography{bib/citations,bib/convsearch,bib/ourbib,bib/ref,bib/ref2,bib/preethabib,bib/emg,bib/nlpa-proposal,bib/paper.bib,bib/recent-edu-hci}  










\end{document}